\documentclass[10pt,a4paper]{article}

\usepackage{jstyle}
\usepackage{tikz}
\usepackage{amsthm,mathrsfs,stmaryrd,bm}
\usepackage[vcentermath]{youngtab}
\usepackage{simplewick}
\usepackage{framed,pifont}
\usepackage[]{xcolor}

\usepackage{graphics} 
\usepackage{graphicx}
\usepackage{epic}
\usepackage{eepic}

\usepackage{feynmp}
\usepackage[retainorgcmds]{IEEEtrantools}
\usepackage{tensor}

\usepackage{verbatim}
\usepackage{latexsym}
\usepackage{amsmath}
\usepackage{wasysym}
\usepackage{amsfonts}
\usepackage{upgreek}
\usepackage{dsfont}
\usepackage{epsf}
\usepackage{epsfig}
\usepackage{amssymb,bm,epsf,epsfig}

\usepackage{color}
\definecolor{rougef}{rgb}{0.56,0,0}
\definecolor{vertf}{rgb}{0,0.5,0}
\definecolor{bleuf}{rgb}{0,0,0.8}
\definecolor{violetf}{rgb}{0.5,0,0.5}

\usepackage[vcentermath]{youngtab}

 \csname
@addtoreset\endcsname{equation}{section}

\def\3s{{s \choose 3}}
\def\4s{{s \choose 4}}
\def\5s{{s \choose 5}}
\def\6s{{s \choose 6}}

\def\12{\dfrac{1}{2}}
\def\fr{\frac}
\def\ft{\footnote}

\def\nn{\nonumber}

\def\2{\ell_2}

\def\pr{\partial}

\def\prd{\partial \cdot}

\def\nn{\nonumber}

\def\be{\begin{equation}}
\def\ee{\end{equation}}
\def\bea{\begin{eqnarray}}
\def\eea{\end{eqnarray}}
\def\ba{\begin{array}}
\def\ea{\end{array}}
\def\bec{\begin{center}}
\def\ec{\end{center}}

\def\a{\alpha} 
\def\b{\beta}  
\def\g{\gamma} 
\def\G{\Gamma}
\def\d{\delta} 

\def\e{\epsilon}

\def\h{\eta}

\def\k{\kappa}
\def\l{\lambda}

\def\m{\mu}
\def\n{\nu}

\def\r{\rho}
\def\s{\sigma}

\def\vf{\varphi}

\def\cA{{\cal A}}

\def\cD{{\cal D}}
\def\cE{{\cal E}}
\def\cF{{\cal F}}

\def\cH{{\cal H}}

\def\cL{{\cal L}}
\def\cM{{\cal M}}
\def\cN{{\cal N}}
\def\cO{{\cal O}}

\def\cR{{\cal R}}
\def\cS{{\cal S}}

\def\ta{{\tilde{A}}}
\def\tal{{\tilde{\alpha}}}
\def\tD{{\tilde{\cD}}}
\def\tf{{\tilde{F}}}

\author[a]{Pietro Ferrero,}
\author[b, c]{Dario Francia}

\affiliation[a]{Mathematical Institute, University of Oxford, Andrew Wiles Building, Radcliffe Observatory Quarter, Woodstock Road, Oxford, OX2 6GG, UK} 
\affiliation[b]{Museo Storico della Fisica e Centro Studi e Ricerche E. Fermi, Piazza del Viminale 1, I-00184 Roma,
Italy}
\affiliation[c]{Roma Tre University and INFN Roma Tre, via della Vasca Navale, 84 I-00146 Roma, Italy} 
\emailAdd{pietro.ferrero@maths.ox.ac.uk, \\ \hskip 37pt dario.francia@cref.it, dario.francia@roma3.infn.it}

\title{
\huge{On the Lagrangian formulation of the double copy to cubic order}}

\abstract{We investigate the Lagrangian formulation of the double-copy correspondence between gauge theories and gravity, up to the cubic order.  Building on the definition of the double-copy field  as a convolution of two vectors, we obtain free gravitational Lagrangians as products of two Yang-Mills Lagrangians, in a form amenable to be easily extended to the massive case. We discuss the origin of these results from tensionless strings and show the existence of gauge fixings that mix the two spin-one sectors and lead to an alternative, especially simple, version of the free Lagrangian. We then construct cubic vertices for the full double-copy multiplet, comprising a graviton, a two-form and a scalar particle, by means of the Noether procedure. Both at the free and at the cubic level the result gets uniquely fixed only upon imposing, on top of gauge invariance, a left-right Lorentz symmetry ruling contraction of indices among double-copy fields. Whereas the outcome nicely matches the cubic interactions of $\mathcal{N}=0$ supergravity, including the gauge-invariant coupling between the scalar particle and the two-form, such a twofold Lorentz symmetry seems to conflict with the perturbative reconstruction of spacetime geometry.}


\begin{document}

\begin{fmffile}{diagram}

\maketitle

\setcounter{tocdepth}{2}
\tableofcontents

\section{Introduction} \label{sec: intro}

The purpose of this work is to explore the Lagrangian counterpart of one of the simplest incarnations of the double copy (DC), connecting spin-one gauge theories and $\cN=0$ supergravity.  Given the dynamical correspondence established between the two frameworks at the level of amplitudes \cite{BCJ_2008, BCJ_2010} as well as its counterpart on the side of classical solutions \cite{BH_and_DC}, it might seem logical to foresee that  (super)gravitational actions should encode somehow the basic data connecting them to their spin one avatars. Of course, such a naive expectation conflicts with the fact that the DC setup relates gauge and gravitational theories in a way that appears to be not just independent from, but even somehow at odds with the redundancies  of the Lagrangian formulation and in particular with the complications emerging when computing amplitudes via Feynman diagrams.\ft{For a recent review and extensive references on the various aspects of the DC see \cite{Bern_review}. See also \cite{Carrasco:2015iwa,White:2017mwc,Bern:2019prr,Borsten:2020bgv}.}. 

Still, the clarification of the status of the DC off shell might be of interest for a number of reasons. To begin with it is expectable, although by no means guaranteed, that it should contribute to better assess the quantum aspects of the duality, providing a path aiming to firmly establish the validity of the DC at loop level, so far still conjectural although robustly supported \cite{BCJ_2010,Du:2012mt,Bern:2013yya,Bern:2017ucb,Bern:2018jmv}.\ft{At the tree level the DC has been proved from several perspectives \cite{Bern_Lagrangian,Feng:2010my,BjerrumBohr:2010hn,Mafra:2011kj,Mizera:2019blq}.} In addition, an off-shell, gauge invariant formulation of the DC could be useful in the context of classical solutions, in themselves typically confined to specific choices of frames and gauge fixings. Furthermore, at the conceptual level, one would like to get some clues on the meaning of the DC  in terms of the geometrical underpinnings of Yang-Mills and gravitational theories. Indeed, the affinities between gauge and gravity theories were most often identified on account of relations between their underlying local symmetries: from the formulation of Einstein's gravity as the gauge theory of a properly identified global group, to the perturbative construction of vertices as  deformations of the Maxwell and Fierz-Pauli actions, and, not last, to holographic dualities. 

If compared to other developments in the subject, however,  the Lagrangian counterpart of the DC was less explored so far.  To date there were only a few explicit attempts: after the precursory work \cite{Bern:1999ji}, first to propose a field-theoretical elaboration of the KLT results \cite{KLT}, an early investigation of the off-shell counterpart of the DC was performed in \cite{Bern_Lagrangian} by Bern, Dennen, Huang and Kiermaier. They proposed an extension of the Yang-Mills Lagrangian, devoid of longitudinal terms due to partial gauge fixing and augmented by suitable nonlocal terms: although the additional vertices in the resulting construction sum up to zero due to the Jacobi identity, still their presence proves to be instrumental to provide amplitudes directly in the color-kinematics dual form. Upon trading non localities for auxiliary fields and after defining an appropriate double-copy dictionary, the authors of \cite{Bern_Lagrangian} were  able to derive a gravitational Lagrangian valid up to five points. Whereas the relation between gauge transformations and diffeomorphisms was not discussed in \cite{Bern_Lagrangian}, the vantage point of symmetries was explicitly taken by  Anastasiou, Borsten, Duff, Hughes and Nagy in \cite{YMorigin}. There they recovered the linearised local symmetries of the \mbox{(super)gravity} multiplet from their spin-one counterparts, employing a suitable definition of gravitational fields in terms of off-shell vectors. Drawing from these ideas, this version of the double copy was extended to the linearised equations of motion in \cite{Comments_DC}. An off-shell dictionary for the BRST ghosts of $\mathcal{N}=0$ supergravity was then considered in \cite{Anastasiou:2018rdx}, still at the linearised level, while the cubic Einstein-Hilbert vertices were recovered in \cite{Borsten:2020xbt}, using a Lagrangian double-copy prescription inspired by that of \cite{Bern_Lagrangian}. A possible pathway to a BRST double copy covering both tree and loop level was recently proposed in \cite{Borsten:2020zgj}. Other attempts aimed at an off-shell understanding of the DC can be found in \cite{Monteiro:2011pc,BjerrumBohr:2012mg,Tolotti:2013caa,Monteiro:2013rya,Cheung:2016prv,Chen:2019ywi}. A systematic exploration of which supergravities qualify as double copies was performed in \cite{Anastasiou:2016csv,Anastasiou:2017nsz,Anastasiou:2017taf}.

Our starting point is the definition of the (extended) gravitational field, or DC field, as a convolution product between two Yang-Mills fields  given in \cite{YMorigin, Comments_DC}:\footnote{See also \cite{Siegel95} for precursory ideas.}
\be \label{defstar}
 H_{\mu\nu}(x)\, = \,  \big[A_{\mu} \, \star \, \tilde{A}_{\nu}\big]\, (x)\, ,
\ee
where the notation is explained in Appendix \ref{sec:notation}. The DC field $H_{\mu\nu}$ defines a reducible representation of $GL (D)$ and qualifies as a natural candidate for an off-shell description of the $\cN = 0$ supergravity multiplet, comprising a graviton, a Kalb-Ramond field and a dilaton.  The main advantage of \eqref{defstar} is that it explicitly connects the linearised  transformations of the  DC field to the Abelian gauge symmetries of the vectors $A_{\mu}$ and $ \tilde{A}_{\nu}$. Building on \eqref{defstar}  we obtain three sets of results. 

In Section \ref{sec:quadratic} we construct  quadratic Lagrangians for $H_{\mu \nu}$ whose equations propagate the full double-copy multiplet. The simplest one,
\be \label{DClagrML_intro}
\cL_{ML}\, = \, \frac{1}{2}\, H^{\m\n} \big( \h_{\r\m} \, \h_{\n\s} \, \Box - \h_{\r\m} \, \partial_{\s} \, \partial_{\n} - \h_{\s\n} \, \partial_{\r} \, \partial_{\m} \big) H^{\r\s}\, ,
\ee
possesses a  gauge symmetry requiring a transversality condition on the parameters. The latter implements a linear dependence between the two gauge sectors of the DC that can be held responsible for the matching of off-shell degrees of freedom between gauge and gravitational sectors. We then show how such a linear dependence between the gauge sectors can be dispensed with by including an additional, unphysical field devoid of a direct interpretation in a strict DC setup, while still keeping locality of the result.  Differently, in a fully gauge-invariant theory formulated in terms of the DC field only the Lagrangian has to be non local and displays most clearly its DC structure:
\be \label{L=FFintro}
\begin{split}
  \cL \, & =  \, \12 \, \cR_{\mu\nu\rho\sigma} \, \fr{1}{\Box}\, \cR^{\mu\nu\rho\sigma} \\
  & = \, \frac{1}{8}(F_{\, \m\a}\star\tf_{\, \n\b})\frac{1}{\Box}(F^{\, \m\a}\star\tf^{\, \n\b}),
\end{split}  
\ee
where $\cR_{\mu\nu\rho\sigma}$ is the linearised field strength of $H_{\mu\nu}$, while $F_{\, \m\n}$ and $\tf_{\, \r\s}$ denote the Yang-Mills curvatures of the corresponding vectors. The nonlocality in \eqref{L=FFintro} has a physical meaning, stemming from the need to covariantly encode the gauge-invariant scalar degree of freedom of the extended gravitational multiplet. At the formal level, it may be understood as the result of integrating out  the unphysical field from the local formulation with unconstrained gauge invariance. Amusingly, the DC Lagrangian \eqref{L=FFintro} admits a direct  massive deformation by the addition of a Proca-like term quadratic in the DC field. 

Let us stress that the form of \eqref{L=FFintro} is not fixed by gauge invariance alone. As we show in Appendix \ref{appendixB}, among the possible gauge-invariant quadratic forms, \eqref{L=FFintro} gets selected by imposing an additional left-right Lorentz symmetry ruling contraction of indices, whose role in the DC was noticed in \cite{Bern:1999ji, Siegel:1993bj,Siegel95,Hohm:2011dz,Cheung:2016say}. Interestingly, both \eqref{L=FFintro} as well as its local counterparts can be interpreted as the rank-two representatives of the class of higher-spin actions described in  \cite{Onthegeometry, ST, FT, triplets, ML}, emerging from tensionless  free strings. At the technical level, the simplicity of the actions \eqref{DClagrML_intro} and \eqref{L=FFintro} derives from the absence of traces of $H_{\m\n}$. With hindsight the latter is to be expected in the DC context, given that the product of the two Lorentz-invariant Yang-Mills factors cannot generate contraction of indices within the same DC field.

We then extend our construction one step beyond the free theory building in Section \ref{sec:cubic} cubic self-interactions for $H_{\m \n}$ by means of the Noether procedure, that we recall in Appendix \ref{noetherAlg}.  Consistency with the free gauge symmetry  fixes the structure of the cubic vertex only up to one free parameter, on top of the overall coupling. This is slightly atypical from the perspective of the Noether procedure and reflects the possibility of consistent couplings involving the scalar mode and the Kalb-Ramond field that are not ruled by gauge symmetry. Again, by imposing that also the cubic vertex respects the twofold Lorentz symmetry acting separately on left and right indices, we fix uniquely the form of the cubic self-interactions of $H_{\m \n}$.  The resulting couplings match the off-shell cubic vertices of the $\mathcal{N}=0$ supergravity Lagrangian, up to field redefinitions, including the Abelian-invariant interaction between the scalar and the Kalb-Ramond field that is not constrained by the Noether procedure. In this sense our result extends  \cite{Bern_Lagrangian} to include the longitudinal parts of the cubic vertex that do not contribute to the amplitude, but that are needed to ensure full off-shell gauge invariance.  

In Section \ref{sec:deformation} we compute the first nonlinear correction to the free gauge transformation of $H_{\mu\nu}$ and start to explore the relation between DC and geometry. We first observe that, in order for both symmetric and antisymmetric parts of $\d H_{\m\n}$ to correspond to the Lie derivative of a rank-two symmetric tensor and of a two-form, respectively, a field redefinition is needed. The latter, however, breaks the left-right Lorentz symmetry that was instrumental both in order to obtain the free Lagrangian in the form \eqref{L=FFintro} and to match the couplings of $\cN=0$ supergravity. Moreover, the same redefinition does not suffice to reproduce the action of the Lie derivative on the scalar field encoded in the trace of the transverse part of $H_{\m\n}$. Rather, in order to reproduce the correct properties of a scalar field we redefine the latter as a suitable nonlinear function of $H_{\mu\nu}$, to be computed perturbatively. We solve the correspondent system of equations to second order in $H_{\mu\nu}$. The result has a clear interpretation from the geometrical viewpoint as a convolution of the Green function of the Laplace-Beltrami operator with the Ricci scalar of a manifold with metric $g_{\mu\nu}=\eta_{\mu\nu}+H^S_{\mu\nu}$ and Levi-Civita connection, where $H^S_{\mu\nu}$ denotes the symmetric part of the DC field.  

Our analysis seems to highlight the existence of an intrinsic tension between the DC setup and Riemannian geometry, in the sense that one can make one side manifest only at the expense of somehow obscuring the other. At the geometrical level, in particular, the relation between gauge symmetries and diffeomorphisms seem to go beyond the literal ``double'' copy paradigm in the sense that, in order for diffeomorphisms to act in the standard fashion, each physical field should be probably interpreted as a power series in the DC field $H_{\m\n}$, and thus, ultimately, in the vectors $A_{\m}$ and $\tilde{A}_{\m}$.

\section{Free Lagrangians for the double copy} \label{sec:quadratic}

In this section we build the free action for the $\cN = 0$ supergravity multiplet as a suitable square of free Yang-Mills Lagrangians and show how to introduce a consistent massive deformation.  Here we will be concerned only with the classical, tree-level DC and for this reason we shall not discuss the role of  ghosts.

\subsection{The double copy field} \label{sec:lagr}  

The  DC field $H_{\m\n}$ together with its gauge transformation are defined as follows \cite{YMorigin}:
\begin{align}
H_{\, \m \n} \, &= \,  \big[A^{a}_{\m} \circ \Phi^{-1}_{aa'} \circ \tilde{A}^{a'}_{\n}\big]\, := \, A_{\m}\,\star\, \ta_{\n},  \label{H} \\
\d_0 H_{\m\n}&= \, \pr_{\mu}\, \a_{\nu} \, + \, \pr_{\nu} \, \tal_{\m},  \label{deltaH}
\end{align} 
where $\circ$ is a convolution product, defined in Appendix \ref{sec:notation}, while $A_{\, \m}$, $\tilde{A}_{\, \n}$ are two Yang-Mills fields in the adjoint representation of two gauge groups $G_L$ and $G_R$, where the suffixes $L$ and $R$ denote left and right factors, respectively. Eq. \eqref{H} also provides the definition of the $\star-$product, while the suffix in $\d_0$ stresses the fact that \eqref{deltaH} defines the local symmetry of the free theory. The field $\Phi_{aa'}$ is a scalar field in the biadjoint representation, whose physical meaning was investigated both for scattering amplitudes \cite{Cachazo:2013hca,Cachazo:2013iea,Cachazo:2014xea} and in the context of classical solutions \cite{White:2016jzc,Goldberger:2017frp,DeSmet:2017rve,Bahjat-Abbas:2018vgo,Luna:2020adi}.  $\Phi^{-1}_{aa'}$ denotes its convolution inverse,
\be
  [\Phi^{-1}_{aa'}\, \circ\, \Phi_{bb'}]\, (x)\, =\, \d_{ab}\, \d_{a'b'}\, \d^{\, (D)}(x)\, ,
\ee
here used  in particular to the goal of building a color singlet. The definition \eqref{H} of the DC field $H_{\mu\nu}$ has proved to be useful in several contexts, including linearised supergravity \cite{Cardoso:2016ngt,Cardoso:2016amd,Borsten:2017jpt} as well as instances of DC on non-trivial backgrounds \cite{Borsten:2019prq}. 

As a tensor, $H_{\, \m \n}$ encodes the rank-two symmetric and antisymmetric irreps of $GL (D)$, that in terms of vector components are given by
\be \label{SA}
\begin{split}
H^S_{\, \m \n} \, &= \, \12(A_{\m}\,\star\, \ta_{\n} \, + \, A_{\n}\,\star\, \ta_{\m}), \\
H^A_{\, \m \n} \, &=\, \12(A_{\m}\,\star\, \ta_{\n} \, - \, A_{\n}\,\star\, \ta_{\m})\, ,
\end{split}
\ee
while the gauge parameters $\a_{\m}$ and $\tilde{\a}_{\m}$ are defined as follows: 
\be
\begin{split} \label{eq:gaugepar}
\a_{\, \m}  \, =\, \e \star\tilde{A}_{\mu}\, , \\
\tilde{\a}_{\mu} \, = \, A_{\mu}\star  \tilde{\e}\, .
\end{split}
\ee
The tensors  in \eqref{SA} transform as an off-shell graviton and an off-shell two-form gauge field,  whereas in order to achieve a full description of the DC multiplet we need to also enforce the propagation of a massless scalar, consistently with the product of two vector irreps of  $O(n)$:
\be \label{YTDC}
\young(\hfil) \, \otimes \, \young(\hfil) \, = \, \young(\hfil\hfil) \, \oplus \, \young(\hfil,\hfil) \, \oplus \, \bullet\, .
\ee
The scalar has to sit in the trace of the DC field $H_{\m\n}$ in some non-trivial manner, though, given that at face value $H^{\a}{}_{\a}$ could be gauged away exploiting \eqref{deltaH}, as it happens for the trace of the ordinary graviton. In this sense, the non-trivial aspect of the free Lagrangian description of the DC multiplet concerns the propagation of the massless scalar. 

The same issue presents itself for the DC of {\it massive} vectors, for which one may still make use of the DC field \eqref{H} without implementing the local transformations \eqref{deltaH}: whereas its on-shell content is still described by \eqref{YTDC},\ft{The matching of dimensions between l.h.s. and r.h.s. of \eqref{YTDC} reads
$$
n^2 \, = \, \12\, (n-1)\, (n+2) \, + \, \12 \, n \, (n-1) \, + \, 1 \, .
$$ One recovers the matching of d.o.f. among the corresponding set of particles upon substituting $n$ with $D-2$ for massless particles and with $D-1$ for massive ones.} its off-shell implementation starting from the DC field \eqref{H} should provide equations of motion suitable to imply the correct mass-shell condition also for the trace of $H_{\m\n}$. This rules out, for instance, the possibility to employ the massive Fierz-Pauli equations for $H^S_{\m\n}$, since they set to zero the trace of the graviton. 

For the massless case, denoting with $\vf$ such a  gauge invariant scalar, with $h_{\mu\nu}$ the symmetric graviton field and with $B_{\mu\nu}$ the antisymmetric two-form, the change of basis $H_{\m\n} \longrightarrow (h_{\m\n}, B_{\m \n}, \vf)$ is defined as follows 
\be \label{hBphi}
\begin{split}
h_{\mu\nu} &=\, H^S_{\mu\nu} \, - \, \gamma\, \eta_{\m\n} \, \vf,\\
B_{\mu\nu} &=\, H^A_{\mu\nu},\\
\vf & =\, H \, - \, \frac{\pr\cdot\pr\cdot H}{\Box},
\end{split}
\ee
where in particular  the physical scalar $\vf$ lies in the trace of the transverse part of $H_{\mu\nu}$ whose covariant definition entails the projection encoded in \eqref{hBphi}, since this is the only possibility to covariantly identify a gauge-invariant scalar within $H_{\mu\nu}$.\ft{The need for a nonlocal projection in the definition of the scalar degree of freedom was noticed in \cite{Comments_DC} and \cite{pietro_tesi}. See also \cite{Luna:2016hge} for a gauge-fixed version of \eqref{hBphi}.} By gauge symmetry alone, the trace of the graviton might mix with $\vf$ through a parameter $\g$  whose value is in principle arbitrary. Let us notice, however, that in correspondence of the value 
\be
\g = \fr{1}{D-2}\, 
\ee
the Lorenz gauge on both Yang-Mills fields implies the de Donder gauge for the graviton $h_{\m\n}$. As we will see in Section \ref{sec:N=0}  this special value of  $\g$ also plays an important role in the identification of the fields of the DC multiplet with those appearing in the action of $\cN = 0$ supergravity. It is relevant to mention that the definition of the scalar field in \eqref{hBphi} is tailored to the free theory and may need to be corrected by higher-order terms at the interacting level, as we will see in Section \ref{sec:deformation}. 

One can build a gauge-invariant field strength for $H_{\m\n}$,\ft{We choose a different normalisation with respect to \cite{YMorigin}.}
\be \label{curvature}
\cR_{\m\n\r\s} \, :=\, -\, \12 F_{\m\n}\, \star\, \tf_{\r\s},
\ee
where $F_{\m\n}$ and $\tf_{\r\s}$ are the linearised field strengths for the vectors $A_{\m}$ and $\ta_{\n}$, respectively. As an element of $GL(D)$, $\cR_{\m\n\r\s}$ decomposes  as follows
\be \label{curvaturetableaux}
\cR_{\m\n\r\s} \, =\,\young(\hfil,\hfil) \, \otimes \, \young(\hfil,\hfil) \, = \, \young(\hfil\hfil,\hfil\hfil) \, \oplus \, \young(\hfil\hfil,\hfil,\hfil) \, \oplus \, \young(\hfil,\hfil,\hfil,\hfil) \, ,
\ee
thus suggesting that the geometry associated to $H_{\m\n}$ be that of a manifold with torsion.

\subsection{Free Lagrangians} \label{sec:lagr}  

We would like to build a gauge-invariant action for the DC field $H_{\mu\nu}$ whose equations of motion propagate the full DC multiplet comprising graviton, two-form and scalar particle. We will present three equivalent and strictly related solutions.

\subsubsection{Local solution}

The problem under scrutiny is a special case of the task concerning the construction of a Lagrangian for the maximal $SO(D-2)-$multiplet contained in a given reducible representation of $GL(D).$\ft{More explicitly, if possibly pedantic: consider the irreps of $GL(D)$ contained in the given original $GL(D)-$reducible representation and branch them in the corresponding $SO(D)$ irreps. Each of these irreps, upon enforcing the proper equations and gauge conditions, provides the degrees of freedom of a corresponding $SO(D-2)$ irrep, {\it i.e.} of a massless particle. The set of all these particles is what we refer to as the maximal $SO(D-2)-$multiplet contained in a given $GL(D)-$reducible tensor.} The general solution was found in \cite{ML} for arbitrary $GL(D)-$reps. In the case at stake, corresponding to the $(1,1)-$reducible representation of $GL(D$), it was shown in \cite{ML} that the equations of motion
\be \label{DCeomML}
\Box \, H^{\m\n} \, - \, \partial_{\m} \, \partial^{\r} \, H_{\r}{}^{\n} \, -\, \partial_{\n} \, \partial^{\s} \, H^{\m}{}_{\s} \, = \, 0 \, ,
\ee
stemming from the Maxwell-like Lagrangian
\be \label{DClagrML}
\cL_{ML}\, = \, \frac{1}{2}\, H^{\m\n} \big( \h_{\r\m} \, \h_{\n\s} \, \Box - \h_{\r\m} \, \partial_{\s} \, \partial_{\n} - \h_{\s\n} \, \partial_{\r} \, \partial_{\m} \big) H^{\r\s}\, ,
\ee
do describe the free propagation of a graviton, a two-form and a massless scalar.  The variation of  \eqref{DClagrML} under \eqref{deltaH},
\be \label{varL0}
\d_0 \, \cL_{ML} \, = \, - \, \pr^{\m} (\a_{\m} + \tal_{\m}) \, \prd \prd H \, ,
\ee
shows that gauge invariance of the theory relies on the transversality constraint
\be \label{tdiff}
\pr^{\, \m} \, (\a_{\m} \, + \, \tal_{\, \m}) \, = \, 0 \, ,
\ee
which provides a generalisation of the condition constraining diffeomorphisms in the various incarnations of unimodular gravity (see {\it e.g.} \cite{uni_einst, uni_old, uni_new}) as well as of its higher-spin extensions \cite{Skvortsov:2007kz,ML, ML_cubic}. Under \eqref{tdiff} the trace and the double divergence of $H^{\m\n}$ are separately gauge invariant. 

In the context of the field theoretical elaboration of the DC it was noticed that two issues typically arise off-shell \cite{Comments_DC,Anastasiou:2018rdx,Borsten:2020xbt}: on the one hand, the number of degrees of freedom of the DC field does not seem to match that of the extended gravitational multiplet, since for instance in $D=4$, upon discarding those gauge components  that can be fixed without making use of the equations of motion, one would be left with $3 \times 3$ components for the DC field, against $6 + 3 + 1$ for the off-shell system given by graviton, Kalb-Ramond field and dilaton. On the other hand, when interpreted in DC terms, the source for the scalar field naturally mixes with the trace of the stress-energy tensor sourcing the graviton equations of motion, thus posing an issue of relative dependence of the two types of couplings.

It seems to us that these issues can be dealt with in the framework of the  theory defined by \eqref{DClagrML}. Indeed,  the constraint \eqref{tdiff} implies a linear dependence between the two gauge sectors that in particular reduces by one scalar component the possibility of performing independent gauge transformations off shell, thus restoring the matching between the expected DC components and their vector seeds.  Moreover, it is relevant to recall that for unimodular gravity the actual source of the gravitational field effectively is the traceless part of the stress-energy tensor, since the variational derivative with respect to metrics with fixed determinant selects the traceless component of the Einstein tensor. There is no missing information with respect to the full sourced Einstein-Hilbert equations, though, since the two setups are known to be  classically equivalent (barring the different role played by the cosmological constant). In this respect, for the symmetric part of $H_{\m\n}$, \eqref{DClagrML} is tantamount to a version of unimodular gravity where the trace of the symmetric tensor is not just frozen, as it could consistently chosen to be, nor is it pure gauge as for the Fierz-Pauli case, rather it represents an independent physical scalar field. It is then clear that, if one were to source \eqref{DCeomML} with a traceful tensor, the trace of the latter would unambiguously source  the scalar field alone encoded in the trace of $H_{\m\n}$. 

Reducible multiplets of particles encoded in a single field are known to be related to the tensionless limit of free string theory, as originally discussed in \cite{triplets_old1, triplets_old2, triplets_old3} and later in \cite{Onthegeometry, ST, FT, triplets, FTbcfw, Lowspinmodels, Gen_connections}. The corresponding BRST construction produces decoupled massless Lagrangians consistent in any dimension, involving ``triplets'' of fields: one physical reducible tensor field together with a number of additional unphysical fields (in particular two additional fields for each physical field in the first Regge trajectory, whence their name), instrumental to enforce unconstrained gauge invariance. In the case of interest for us the relevant member of the tensionless Lagrangian is a simple modification of \eqref{DClagrML},
\be \label{tripletnosymm}
\cL = \frac{1}{2}\, H^{\a\b} \big( \h_{\a\m} \, \h_{\b\n} \, \Box - \h_{\a\m} \, \partial_{\b} \, \partial_{\n} - \h_{\b\n} \, \partial_{\a} \, \partial_{\m} \big) H^{\m\n} + 2 \, H^{\m\n} \, \partial_{\m} \, \partial_{\n} \, D - 2 \, D \, \Box\, D \, ,
\ee
which is indeed fully invariant under \eqref{deltaH} provided that the additional scalar field $D$ transforms as 
\be \label{varD}
\d D \, = \, \pr^{\, \m} \, \Big(\fr{\a_{\, \m} \, + \, \tilde{\a}_{\, \m}}{2}\Big) \,, 
\ee
and where the third field of the triplet has been eliminated by means of its algebraic equations of motion. Let us observe that, for this local Lagrangian with unconstrained gauge invariance, the role of the additional scalar off-shell component is played by the field $D$.

The analysis of the particle content of \eqref{DClagrML} and \eqref{tripletnosymm} is essentially identical and can be easily discussed.
For instance, starting from the equations of  \eqref{tripletnosymm} 
\be \label{eomHD}
\begin{split}
E^H_{\m\n}&:= \, \Box \, H_{\m\n} \, - \, \pr_{\m} \, \pr^{\a}H_{\a\n} \, - \, \pr_{\n} \, \pr^{\a}H_{\m\a} \, + \, 2 \, \pr_{\m} \, \pr_{\n} \, D \,  = \, 0\, ,\\
E^D & := \, \Box \, D \, - \, \12 \, \prd \prd H \, = \, 0 \, ,
\end{split}
\ee 
one can set $D=0$ by a gauge fixing and then separate symmetric and antisymmetric components of the first of \eqref{eomHD}, thus getting two independent equations for $H^S_{\m\n}$ and for $H^A_{\m\n} = B_{\m\n}$:
\begin{align}
& \Box \, H^S_{\m\n} \, - \, \pr_{\m} \, \prd H^S_{\n} \, - \, \pr_{\n} \, \prd H^S_{\m} \,= \, 0, \label{eomHS} \\
& \Box \, B_{\m\n} \, - \, \pr_{\m} \, \pr^{\a}B_{\a\n} \, + \, \pr_{\n} \, \pr^{\a}B_{\a\m} \, = \, 0.  \label{eomHA} \, 
\end{align}
If one were to start with \eqref{DClagrML}, on the other hand, one would directly get \eqref{eomHS} and \eqref{eomHA}. In \eqref{eomHS} the double divergence of $H^S_{\m\n}$ vanishes on shell, consistently with the equation for $D$ in the formulation starting from \eqref{tripletnosymm}, while $\pr^{\, \m} H^S_{\m\n}$
can be gauge fixed to zero  solving for $\Box (\a_{\m} + \tal_{\m})$, since  the transversality of the parameter matches the transversality of $\pr^{\m}  H^S_{\m\n}$, to finally provide the system 
\be
\begin{split}
&\Box \, H^S_{\m\n} \, = \, 0 \, ,\\
& \pr^{\, \a} H^S_{\a \n} \, = \, 0 \, ,
\end{split}
\ee
describing indeed the free propagation of a graviton and a massless scalar, the latter encoded in the here gauge-invariant trace of $H^S_{\m\n}$. Equation \eqref{eomHA}, in its turn,  describes the propagation of a two-form, with gauge transformations given by 
\be
\d_0 B_{\m\n}\, = \, \pr_{\mu}\, \e_{\nu} \, - \, \pr_{\nu} \, \e_{\m},
\ee
where 
\be
\e_{\m} \, = \, \12(\a_{\m} - \tal_{\m})\, ,
\ee
and corresponding gauge for gauge invariance under $\d \e_{\m} = \pr_{\m} \r$.  

\subsubsection{Nonlocal solution}

From the perspective of the DC one may observe that the theory described by \eqref{tripletnosymm}  requires an additional field with respect to $H_{\m\n}$, with no obvious meaning in terms of component spin-one fields. The Lagrangian \eqref{DClagrML}, in its turn, is fully expressed in terms of the DC field, but it requires to enforce the  restriction \eqref{tdiff} on the gauge parameters $\a_{\m}$ and $\tal_{\m}$, which effectively implies a linear dependence of the divergences of $A_{\m}$ and $\tilde{A}_{\m}$.  

It is actually possible to construct a gauge-invariant Lagrangian devoid of these issues, but the resulting expression turns out to be nonlocal. The result is still rewarding, however, as it makes the DC structure apparent. One possibility to get the desired result is to integrate out the field $D$ from \eqref{tripletnosymm}, so as to also keep the origin of the resulting nonlocalities transparent. The calculation is straightforward and gives:
\be \label{Lnonloc}
\cL_{NL} \, = \, \12 H^{\a\b}\big\{ \Box \eta_{\a\mu}\eta_{\b\nu}-\pr_{\a}\pr_{\mu}\eta_{\b\nu}
  -\pr_{\b}\pr_{\nu}\eta_{\a\mu}+\pr_{\a}\pr_{\b}\frac{1}{\Box} \pr_{\mu}\pr_{\nu}
  \big\}H^{\mu\nu}\,.
\ee
Integrating by parts and  bearing in mind the definition \eqref{curvature} of the DC field strength one finds that the free Lagrangian \eqref{Lnonloc} admits the rewritings
\be 
\begin{split}
\cL_{NL} \, &=\, \12 \, \cR_{\mu\nu\rho\sigma} \, \fr{1}{\Box}\, \cR^{\mu\nu\rho\sigma} \, \label{L=RR} \\
                      &= \, \frac{1}{8}\, (F_{\m\a}\, \star\, \tf_{\n\b})\, \frac{1}{\Box}\, (F^{\m\a}\, \star\, \tf^{\n\b}) \, ,
\end{split}
\ee
with $\cR_{\mu\nu\rho\sigma} $ defined in \eqref{curvature}. It can be useful to write $\cR_{\mu\nu\rho\sigma}$ in terms of $H^S_{\m \n}$ and $B_{\m \n}$:
\be \label{R=EH+KR}
 \begin{split}
 \cR_{\m \n \r \s} & =  \12\, \big\{\pr_{\n}\pr_{\r}H^S_{\m \s} + \pr_{\m} \pr_{\s} H^S_{\n\r} - \pr_{\n}\pr_{\s} H^S_{\m \r} - \pr_{\m} \pr_{\r} H^S_{\n \s}\big\} \\ 
& + \12\, \big\{\pr_{\nu}\pr_{\rho}B_{\mu\sigma}+\pr_{\mu}\pr_{\sigma}B_{\nu\rho}-\pr_{\nu}\pr_{\sigma}B_{\mu\rho}-\pr_{\mu}\pr_{\rho}B_{\nu\sigma}\big\},
 \end{split}
\ee 
where in the first row one can recognise the linearised Riemann tensor $R^S_{\m \n \r \s}$ of a manifold with Levi-Civita connection and metric $g_{\m \n } = \h_{\m \n } + H^S_{\m \n}$, while in the second row the field strength of the two-form, 
\be
\cH_{\m\n\r} = \pr_{\m} B_{\n\r} + \pr_{\r} B_{\m\n}  + \pr_{\n} B_{\r\m}\, ,
\ee
appears in the combination $\pr_{\m} \, \cH_{\n\r\s} \, - \, \pr_{\n} \, \cH_{\m\r\s}$.
Substituting \eqref{R=EH+KR} in \eqref{L=RR} one obtains
\be
\cL_{NL} \, =\, \12 \, R^S_{\mu\nu\rho\sigma} \, \fr{1}{\Box}\, R^{S\mu\nu\rho\sigma} \, - \, \fr{1}{6} \,  \cH_{\m\n\r} \,  \cH^{\m\n\r}\, ,
\ee
where the first term is the spin-two representative of the class of Lagrangians obtained in \cite{triplets} from the first Regge trajectory of free strings collapsed to zero tension, while the second term is the action of a two-form gauge field, whose origin can be traced back to  the presence of a torsion, identified as the field strength of the two form up to a sign:
\be
T_{\, \m \r \s} \, = \, - \, \cH_{\, \m \r \s} \, .
\ee
The nonlocal equations of motion obtained from \eqref{Lnonloc}
\be \label{DCeomNL}
\Box \, H^{\mu\nu} \, -\, \pr^{\mu} \pr_{\a} \, H^{\a \nu}\, - \, \pr^{\nu} \pr_{\b}\, H^{\mu\b}\, +\, 
\pr^{\m}\, \pr^{\n} \frac{1}{\Box} \pr_{\a} \, \pr_{\b} \, H^{\a\b}\, =\, 0\, ,
\ee
are manifestly equivalent to their local counterparts \eqref{eomHD} or \eqref{DCeomML}, as the Lagrangian \eqref{Lnonloc} was obtained integrating away from \eqref{tripletnosymm} a field that does not carry physical degrees of freedom. Alternatively, one can observe that they reduce to \eqref{DCeomML} by a partial gauge fixing.

It may be interesting to notice that, following \cite{Lowspinmodels}, our massless DC Lagrangians \eqref{L=RR} admit a simple extension to massive 
Lagrangians of the Proca-like form
\be  \label{L=RR+m}
\begin{split}
\cL_{NL} \, &=\, \12 \, \cR_{\mu\nu\rho\sigma} \, \fr{1}{\Box}\, \cR^{\mu\nu\rho\sigma} \, - \, \12 \, m^2 \, H_{\m\n}\, H^{\m\n} \\
                      &= \, \frac{1}{8}\, (F_{\m\a}\, \star\, \tf_{\n\b})\, \frac{1}{\Box}\, (F^{\m\a}\, \star\, \tf^{\n\b}) \, - \, \12 \, m^2 \, 
                      (A_{\m}\,\star\, \ta_{\n})\, (A^{\m}\,\star\, \ta^{\n})\, .
\end{split}
\ee
Indeed, the corresponding equations of motion
\be \label{DCeomNLm}
\Box \, H^{\mu\nu} \, -\, \pr^{\mu} \pr_{\a} \, H^{\a \nu}\, - \, \pr^{\nu} \pr_{\b}\, H^{\mu\b}\, +\, 
\pr^{\m}\, \pr^{\n} \frac{1}{\Box} \pr_{\a} \, \pr_{\b} \, H^{\a\b}\, - \, m^2 \, H^{\m\n}\, =\, 0\, ,
\ee
imply
\be \label{divH}
\pr_{\a} \, H^{\a \nu}\, = \, 0 \, = \, \pr_{\b}\, H^{\mu\b}\, ,
\ee
and thus reduce to
\be \label{massive}
(\Box \, - \, m^2) \, H^{\m\n}\, =\, 0\, ,
\ee
that describe the free propagation of massive graviton, two-form and scalar particles, as can be seen upon decomposing $H^{\mu\nu}$ in its symmetric-traceless,  antisymmetric and trace parts.

 Let us mention that, if we were to derive our results anew, without making reference to their origin from tensionless strings, gauge invariance alone would not suffice  to select a unique Lagrangian for $H_{\m\n}$, neither in the local form \eqref{DClagrML} nor in its nonlocal counterpart \eqref{Lnonloc}, so that one may wonder what is intrinsically special of the selected terms. As we detail in Appendix \ref{appendixB} one can see that, among the possible gauge invariant bilinears,  \eqref{DClagrML} and \eqref{L=RR} get selected as the unique Lagrangians (local and nonlocal, respectively) that display a twofold Lorentz symmetry where each index in $H_{\m\n}$ undergoes an independent Lorentz transformation. The relevance of such a twofold Lorentz symmetry, consistent with the factorisation property of the KLT relations, was noticed already in the first attempts to a Lagrangian formulation of the DC in \cite{Bern:1999ji} and more recently stressed in particular in  \cite{Siegel:1993bj,Siegel95,Hohm:2011dz,Cheung:2016say}. As we shall see in the Section \ref{sec:cubic}, it will play a role also in the construction of the cubic vertex.

\subsection{Double-copy analysis of the equations of motion} \label{sec:lagr}  

Here we would like  to comment on the relation between the equations \eqref{DCeomML} and \eqref{DCeomNL} and the corresponding free equations of motion for the vectors $A_{\m}$ and $\tilde{A}_{\m}$. To this end, we first rewrite \eqref{DCeomML} in terms of vectors:
\be \label{eomMLA}
\Box \, (A_{\m}\,\star\, \ta_{\n}) \, - \, \partial_{\m} \, \partial^{\a} \, (A_{\a}\,\star\, \ta_{\n}) \, -\, \partial_{\n} \, \partial^{\a} \, (A_{\m}\,\star\, \ta_{\a}) \, = \, 0 \, .
\ee
It is then easy to see, making use in particular of the property \eqref{nonLei} of the $\star-$product, that if both vectors satisfy their own free equations,  then \eqref{eomMLA}  holds, at least in any gauge implying transversality of one of the vectors. For instance, if  $\Box A_{\m} - \partial_{\m}  \partial^{\a}  A_{\a} = 0$ and we choose a gauge implying $\pr^{\, \a} \ta_{\a} = 0$, then it is immediate to see that \eqref{eomMLA} does vanish. Let us observe that the transversality condition does not need to be imposed directly as the Lorenz gauge, but can follow in other gauges on shell, {\it e.g.} in light-cone coordinates, in a frame where $p_+ \neq 0$, one can choose a gauge s.t. $\ta_+ = 0$ and in that gauge, on shell, one finds $\pr^{\a} \ta_{\a} = 0$. This situation clearly applies to \eqref{DCeomNL} as well, given that if one vector is transverse the nonlocal equations reduce to \eqref{DCeomML}. It may seem possible to ask for a weaker condition: namely to have only one vector  on shell and the second vector in the Lorenz gauge, for the field $H_{\m \n}$ to be on shell, as both \eqref{DCeomML} and \eqref{DCeomNL} would hold. However, the support of the physical solutions of \eqref{eomMLA} is on the light cone $p^2=0$, and this applies to both vectors $A_{\m}$ and $\ta_{\m}$ in the convolution defining $H_{\m\n}$. 

The fact that the equations of motion for the gauge vectors need to be supplemented with gauge conditions in order to grant for the DC multiplet to be on shell was already noticed in  \cite{Comments_DC}. This is in some sense to be expected, given that the Maxwell equations are transverse and thus cannot provide direct information on the divergence of the vectors without supplementing them with some gauge condition. 

The situation is even simpler for the massive equations \eqref{massive}  and \eqref{divH}, that from the DC perspective immediately prove to be equivalent to the Proca equations
\be
\begin{split}
(\Box \, - \, m^2) \, A^{\m}\,  & =\, 0\, , \\
\pr^{\m} \, A_{\m} \, &= \, 0 \, ,
\end{split}
\ee
for both vectors in the DC field.\footnote{Let us stress that in the present construction the two vectors in the DC field are bound to have the same mass, equal to that 
of all the particles in the gravitational multiplet subsumed by \eqref{DCeomNLm}.}

\section{Cubic interactions} \label{sec:cubic}

In extending the free DC Lagrangians \eqref{DClagrML}, \eqref{tripletnosymm} and \eqref{L=RR} to the interacting level, one meets a few obvious issues that it may be worth to briefly recall.

To begin with, trading the linearised Yang-Mills field strength in \eqref{L=RR} with the full one cannot work: the number of derivatives in the interaction vertices would not reproduce the two-derivative interactions of General Relativity and the resulting polynomial theory would not match  the expected non-polynomial vertices of the Einstein-Hilbert formulation.\footnote{An alternative option, that we do not pursue in this paper, would be to try to make contact with polynomial formulations of gravity \cite{Palatini_history, Deser_self, Cheung:2017kzx}.}

The gluonic Lagrangians considered in \cite{Bern_Lagrangian} for the squaring process (see also \cite{Tolotti:2013caa}) involve higher-order, nonlocal vertices to the goal of making the color-kinematics duality explicit. The square of such modified theory is expected to produce a non-polynomial gravity theory. In \cite{Bern_Lagrangian} both non localities and higher-order vertices are dealt with via the introduction of auxiliary fields. However, whereas the nonlocal terms added to the Yang-Mills Lagrangians are effectively zero  due to the color Jacobi identity, it is less clear which role would they play when employed in the squaring process, since  in that context the only remainder of the Jacobi identity at the spin-two level is the full antisymmetrisation of the corresponding gravitational terms. Given the equivalence of the two theories at the amplitude level one may guess that the differences with the Einstein-Hilbert Lagrangian would be confined to sectors involving longitudinal components of the graviton. Still, it would be interesting to explicitly identify the fate of the spin-two nonlocalities inherited from the spin-one, parent theory, in the setup where all contact interactions are kept and no auxiliary fields are introduced.
 
On the technical side, an actual cubic vertex for $H_{\m\n}$ would have the schematic form 
\be
\cS_1 \, =\, \int d^Dx (A_1\star\ta_1)(A_2\star\ta_2)(A_3\star\ta_3)\, ,
\ee
whereas a naive vertex stemming from the inclusion of the non-Abelian part of the Yang-Mills  field strength in \eqref{L=RR}, for instance,  would contain terms of the type \footnote{We neglect color indices, derivatives and structure constants as well as the biadjoint scalar in the $\star$-product, since they do not play any role in this observation.}
\be
\hat{\cS}_1 \, = \, \int d^Dx [(A_1A_2)\star(\ta_1\ta_2)](x)[A_3\star\ta_3](x) \, ,
\ee
where the identification of three DC fields would be somewhat ambiguous.

Thus,  it appears that the extension of the double copy at the interacting level involves additional subtleties that defy an immediate intuitive understanding of the eventual outcome. For these reasons, we shall choose a bottom-up, systematic, approach to construct cubic interactions as consistent deformations of given free Lagrangians, following the Noether procedure.

\subsection{Cubic vertex: TT sector}

Abiding by the Noether procedure, recalled in Appendix \ref{noetherAlg}, as a first step we write an ansatz for the  transverse-traceless (TT) part of the vertex and try to fix the corresponding coefficients upon imposing invariance under the free gauge transformations of $H_{\m\n}$. We assume the vertex to contain two derivatives, given that we aim to reproduce, among other couplings, standard gravitational interactions, and to be local, since we do not want to introduce nonlocalities in the physical sector of the theory. Let us stress that the TT part of the cubic vertex extends both the local theory \eqref{DClagrML} and its nonlocal counterpart \eqref{L=RR}, since they  differ only in the sector involving the double divergence of $H_{\m\n}$.  Moreover, even if we were to start from \eqref{tripletnosymm}, we would not include the field $D$ in the TT vertex, as its gauge transformation \eqref{varD} makes it a natural candidate to enter counterterms involving longitudinal components of $H_{\m\n}$ (or possibly traces, that will not enter anyway our vertices). In order to keep track of equivalences up to integration by parts we employ the cyclic ansatz (see {\it e.g.}  \cite{ML_cubic}), according to which derivative indices always contract with the tensor appearing to the left of the derivative operator.  The general form of the TT vertex is thus the following one
\begin{align}\label{eq:L1HTT}
\begin{split}
   \cL_1^{\text{TT}}&=a_1 H^{\mu\nu}\pr_{\mu}\pr_{\nu}H_{\a\b}H^{\a\b}+
   a_2H^{\mu\nu}\pr_{\mu}\pr_{\nu}H_{\a\b}H^{\b\a}\\ &+
   a_3H^{\mu\nu}\pr_{\mu}H^{\a\b}\pr_{\a}H_{\b\nu}+
   a_4H^{\mu\nu}\pr_{\nu}H^{\a\b}\pr_{\a}H_{\b\mu}+
   a_5H^{\mu\nu}\pr_{\mu}H^{\a\b}\pr_{\b}H_{\a\nu}\\  &+
   a_6H^{\mu\nu}\pr_{\nu}H^{\a\b}\pr_{\b}H_{\a\mu}+
   a_7H^{\mu\nu}\pr_{\mu}H^{\a\b}\pr_{\a}H_{\nu\b}+
   a_8H^{\mu\nu}\pr_{\nu}H^{\a\b}\pr_{\a}H_{\mu\b}\\ 
    &+
   a_9H^{\mu\nu}\pr_{\mu}H^{\a\b}\pr_{\b}H_{\nu\a}+
   a_{10}H^{\mu\nu}\pr_{\nu}H^{\a\b}\pr_{\b}H_{\mu\a}\, ,
\end{split}
 \end{align}
whose gauge invariance  under \eqref{deltaH} holds for a  two-parameter family of solutions:
 \begin{gather}\label{eq:coeffHTT}
   a_5=a_8=a,\qquad a_3=a_4=a_6=a_7=a_9=a_{10}=b, \qquad a_1=a+b, \qquad  a_2=2b.
 \end{gather}
This multiplicity of solutions is  slightly unusual, as the Noether procedure at the TT level typically fixes all the coefficients up to an overall coupling. However, one has to keep in mind that the theory we are constructing also involves a coupling among  the dilaton and the field strength of the two form of the schematic type
\be
\sim \, \vf \, \pr B \, \pr B \, 
\ee
that, being independently gauge invariant, cannot be fixed by the Noether algorithm alone. Actually it is possible to show that this is the only coupling that is not fixed by gauge invariance of the TT vertex.
Nevertheless, this ambiguity disappears if we enforce also at the TT cubic level the twofold Lorentz symmetry that forbids contractions of ``left'' indices with ``right'' indices (see Appendix \ref{appendixB} for more details). Upon discarding all terms that would violate this symmetry one finds the unique TT vertex
\be \label{TTcubic}
\cL_1^{TT}=H^{\mu\nu}\pr_{\mu}\pr_{\nu}H_{\a\b}H^{\a\b}+H^{\mu\nu}\pr_{\mu}H^{\a\b}\pr_{\b}H_{\a\nu}+H^{\mu\nu}\pr_{\nu}H^{\a\b}\pr_{\a}H_{\mu\b}\, ,
\ee
up to the overall coupling. From \eqref{TTcubic} it is  possible to compute the three-point amplitude:
\begin{gather} \label{3point}
\cM_3= \e^{(1)}_{\a\b}\e^{(2)\, \a\b}\e^{(3)}_{\mu\nu}p_2^{\mu}p_2^{\nu} +
 \e^{(1)}_{\mu_1\nu_1}\e^{(2)\, \mu_1\nu_3}\e^{(3)}_{\mu_3\nu_3}p_3^{\nu_1}p_2^{\mu_3} + 
\e^{(1)}_{\mu_1\nu_1}\e^{(3) \mu_3\nu_1}\e^{(3)}_{\mu_3\nu_3}p_3^{\mu_1}p_2^{\nu_3}+\text{\it perms.},
\end{gather}
where $\e_{\mu\nu}$ is the transverse polarisation tensor of $H_{\mu\nu}$. It can be explicitly checked that, upon properly choosing the overall coupling for \eqref{TTcubic}, the amplitude \eqref{3point} matches the simplest, three-point incarnation of the general DC prescription 
\be \label{DC}
\cA_n \, = \, g^{\, n-2} \, \sum_{i\in\G_3} \fr{n_i \, c_i}{\Pi_{\a_i} \, s_{\a_i}} \qquad \longrightarrow \qquad 
\cM_n \, = \, i \, (2 \k)^{\, n-2} \, \sum_{i\in\G_3} \fr{n_i \, \tilde{n}_i}{\Pi_{\a_i} \, s_{\a_i}} \, ,
\ee
where $\cA_n$ and $\cM_n$ represent gauge and $\cN = 0$ supergravity amplitudes, respectively, the former written in the color-kinematics dual explicit form \cite{BCJ_2008, BCJ_2010}. 

In the following we shall build the off-shell completion of \eqref{TTcubic}, whose eventual form depends on the free Lagrangian one starts with. As we will see, in all cases only divergences of $H_{\m\n}$ are needed to this end, consistently with the preservation of the left-right index symmetry that would otherwise be broken by the inclusion of traces. As recalled in Appendix \ref{noetherAlg}, even barring total derivatives and field redefinitions, the Noether procedure may admit several equivalent solutions whenever the free gauge symmetry one starts with is constrained. We will comment about this aspect in Section \ref{sec:deformation}, in connection with the perturbative reconstruction of spacetime geometry. At any rate, the matching between \eqref{3point} and the $n=3$ case of \eqref{DC} makes it clear that any off-shell completion will be consistent with the DC expectations.

\subsection{Local off-shell completion}

To begin with, let us consider \eqref{TTcubic} as the TT cubic deformation of the free local Lagrangian \eqref{DClagrML}. In this case its gauge transformation, keeping \eqref{tdiff} undeformed, can be compensated by the addition of the term
\begin{align} \label{deltaL1ML}
\Delta\cL_1=-\frac{1}{2}\,\prd\prd H\,H_{\a\b}\,H^{\a\b}, 
\end{align}
so that the variation of the full cubic vertex  is
\be\label{varL1ML}
 \d_0 \left(\cL_1^{TT}+\Delta \cL_1\right) \, \approx \, \left(\a^{\b} \,H_{\a\b}\, \pr^{\a} + \tal^{\a} \,H_{\a\b}\, \pr^{\b}\right) \, \prd\prd H \, ,
\ee
where we omitted terms that are manifestly proportional to the equations of motion \eqref{DCeomML} and that as such enter the deformation of the free gauge transformation of $H_{\m\n}$. Actually also the r.h.s. of \eqref{varL1ML}, being proportional to the divergence of the equations of motion, actually is of the same type as the omitted terms, even if it does not contain the d'Alembertian operator. Given the form \eqref{varL0} of the gauge transformation of the free Lagrangian, however, one sees that the r.h.s. of \eqref{varL1ML} can also play a different role. Indeed, as explained in Appendix \ref{noetherAlg}, the Noether procedure for theories with constrained gauge invariance may also involve perturbative deformations of the constraint of the type \eqref{corr}, that in the case at stake would look
\be \label{deform}
\pr^{\m} \, (\a_{\m} + \tal_{\m}) + g\, \cO_1 (\a, \tal, H) \, = \, O\, (g^2) \, ,
\ee
for some linear operator $\cO_1$. Comparing  \eqref{varL0} and  \eqref{varL1ML} it is then clear that one can exploit \eqref{deform} in order to compensate terms in the cubic vertex proportional to the double divergence of $H_{\m\n}$. In force of this possibility, the general solution for the cubic vertex  can be written so as to include an additional parameter $\l$,
\be \label{cubic_local}
\cL_1 (\l)\, =\, \cL_1^{TT}\, - \, \frac{\l}{4}\,\prd\prd H\,H_{\a\b}\,H^{\a\b}\, ,
\ee
such that now the variation looks 
\be \label{deltadef}
\d_0\cL_1 (\l)\approx-\frac{\l-2}{4}\prd\prd H\,\d_0\left( H_{\a\b}\,H^{\a\b} \right)+H^{\a\b}\,\left( \pr_{\a}\prd\prd H\,\a^{\b}+\pr_{\b}\prd\prd H\,\tal^{\a} \right)\, .
\ee
As the form of \eqref{deltadef} suggests, one may choose to compensate the first term by a correction to the constraint of the type \eqref{deform}, namely, taking the overall coupling into account,
\be
\pr^{\m} \, (\a_{\m} + \tal_{\m}) \, = \, -\,g\, \frac{\l-2}{4}\, \d_0\left( H_{\a\b}\,H^{\a\b} \right)\,+\,  O \, (g^2) \, ,
\ee
while the second term, proportional to the divergence of the equations of motion, would contribute directly to $\d_1 H_{\m\n}$. Whereas the separation of terms in \eqref{deltadef} is not forced upon by any consistency condition, and in particular the simplest solution with $\l = 2$ may appear to be preferable at this level, when it comes to computing the first-order correction to the free gauge transformation other choices of $\l$ may acquire geometric significance. We shall comment more on this point in Section \ref{sec:deformation} where we will discuss the computation of $\d_1 H^{\m\n}$. 

Keeping these considerations into account and collecting all terms, the variation of the full cubic vertex under \eqref{deltaH} can be arranged in the form
\begin{align}
\begin{split} \label{deltaL_loc}
\d_0\cL_1\, & =\, M^{\a\b}\,\left[(\a^{\g}+\tal^{\g})\,\pr_{\g}H_{\a\b}-H_{\a\g}\pr^{\g}\a_{\b}-H_{\g\b}\pr^{\g}\tal_{\a}+H_{\a\g}\pr_{\b}\a^{\g}+H_{\g\b}\pr_{\a}\tal^{\g}\right]\\
&-\frac{\lambda-2}{4}\prd\prd H\,\d_0\left( H_{\a\b}\,H^{\a\b} \right)\, ,
\end{split}
\end{align}
where 
\be
M^{\m\n}\, = \, \Box \, H^{\m\n} \, - \, \partial^{\m} \, \partial^{\a} \, H_{\a}{}^{\n} \, -\, \partial^{\n} \, \partial^{\a} \, H^{\m}{}_{\a} \, 
\ee
is the tensor defining the free equations of motion in the present setup. To summarise,  for the local Maxwell-like DC Lagrangian to cubic order we obtained
\be \label{L_loc}
\begin{split}
\cL &= \, \cL_0 \, + \, g\, \cL_1 \, + \, O (g^2) \, \\
      &=  \frac{1}{2}\, H^{\a\b} \big( \h_{\a\m} \, \h_{\b\n} \, \Box - \h_{\a\m} \, \partial_{\b} \, \partial_{\n} - \h_{\b\n} \, \partial_{\a} \, \partial_{\m} \big) H^{\m\n}\, \\
      & + \, g\left( H^{\mu\nu}\pr_{\mu}\pr_{\nu}H_{\a\b}H^{\a\b}+H^{\mu\nu}\pr_{\mu}H^{\a\b}\pr_{\b}H_{\a\nu}+H^{\mu\nu}\pr_{\nu}H^{\a\b}\pr_{\a}H_{\mu\b}\, \right.  \\
      & \left. - \, \frac{\l}{4}\,\prd\prd H\,H_{\a\b}\,H^{\a\b}\right) \, + \, O (g^2) \, .
\end{split}
\ee

One could also consider the possibility to deform to cubic order the free Lagrangian \eqref{tripletnosymm}, involving the auxiliary field $D$. Instead of going through the full Noether procedure anew, let us start from \eqref{TTcubic} as a possible TT cubic deformation of \eqref{tripletnosymm}. Its gauge transformation can be compensated by the addition of the term (see \eqref{deltaL1ML} and \eqref{varL1ML})
\begin{align}
\Delta\tilde{\mathcal{L}}_1 (H, D) =-\tfrac{1}{4} H_{\a\b}\,H^{\a\b}\,\prd\prd H- H_{\a\b}\,\left(\pr^{\a}D\,\pr_{\g}H^{\g\b}+\pr^{\b}D\,\pr_{\g}H^{\a\g}\right)+2 H_{\a\b}\,\pr^{\a}D\,\pr^{\b}D,
\end{align}
in such a way that the variation of the corresponding cubic vertex, 
\begin{align}
\tilde{\cL}_1 (H, D) \, = \, \cL_1^{TT}\, +\, \Delta\tilde{\mathcal{L}}_1 \, (H, D)\, ,
\end{align}
be proportional to the free equations \eqref{eomHD}:
\begin{align}\label{L1_aux_gaugevariation}
\begin{split}
\delta_0 \, \tilde{\cL}_1& = 
E^H_{\alpha\beta}\,\big[-\pr_{\a}D\,\a_{\b}-\pr_{\b}D\,\tal_{\a}-\a^{\g}\,\pr_{\b}H_{\a\g}-\tal^{\g}\pr_{\a}H_{\g\b}-H_{\a\g}\pr^{\g}\a_{\b}-H_{\g\b}\pr^{\g}\tal_{\a}\\
&+(\a+\tal)^{\g}\,\pr_{\g}H_{\a\b}+\tfrac{1}{2}\pr_{\g}(\a+\tal)^{\g}\,H_{\a\b}\big]\\
&+ E^D \,\big[ +\tfrac{1}{2}\delta_0(H_{\a\b}\,H^{\a\b}) - 2 (\a+\tal)^{\g}\pr_{\g}D+2 \a^{\b}\pr^{\a}H_{\a\b}+2\tal^{\a}\pr^{\b}H_{\a\b}\big]\, ,
\end{split}
\end{align}
with the variation of $D$ still given by \eqref{varD}. The possibility to deform the constraint \eqref{tdiff} that we discussed in the previous setup
reflects in an ambiguity in the distribution of the various terms in \eqref{L1_aux_gaugevariation}, some of which may be equivalently interpreted as contributing to the deformation either of $H_{\m\n}$ or of $D$.

\subsection{Nonlocal off-shell completion}
 
An alternative possibility is to start with the free Lagrangian \eqref{L=RR}, where the two spin-ones sectors are fully independent and no auxiliary fields are included. Although the TT sector of the cubic vertex is still given by \eqref{eq:L1HTT}, its off-shell completion  is different due to the different structure of the free equations and to the absence of constraints on the free gauge symmetry. The general strategy of the computation is otherwise quite similar to the one explored in the previous section, so that here we shall simply display the resulting Lagrangian (see \cite{pietro_tesi} for more details):
\be \label{L_utc}
\begin{split}
\cL_{NL} & = \frac{1}{8}\, (F_{\m\a}\, \star\, \tf_{\n\b})\, \frac{1}{\Box}\, (F^{\m\a}\, \star\, \tf^{\n\b}) \\
      & + \, g \left\{
\12 \big[H^{\mu\nu}(\pr_{\mu}\pr_{\nu}H^{\a\b}H_{\a\b}-\pr_{\mu}H^{\a\b}\pr_{\nu}H_{\a\b})+2 H^{\mu\nu}(\pr_{\mu}H^{\a\b}\pr_{\b}H_{\a\nu} \right. \\ 
& \left. + \pr_{\nu}H^{\a\b}\pr_{\a}H_{\mu\b})\big] + H_{\a\b}(2\,\tD^{\a}\cD^{\b}-\tD^{\a}\pr_{\mu}H^{\mu\b}-\cD^{\b}\pr_{\mu}H^{\a\mu})
\right\} \, 
      + \cO (g^2) \, ,
\end{split}
\ee
where we have defined
\begin{align} \label{confession}
\begin{split} 
\cD_{\b}&:=\pr^{\g}H_{\g\b} - \fr{\pr_{\b}}{2 \Box} \prd\prd H,\\
\tilde{\cD}_{\a}&:=\pr^{\g}H_{\a\g} - \fr{\pr_{\a}}{2 \Box} \prd\prd H.
\end{split}
\end{align}
Let us remark that also in this case the cubic term $\cL^1_{NL}$, defined by second and third line in \eqref{L_utc}, displays the same left-right Lorentz symmetry of the free Lagrangian \eqref{L=RR}.  Computing its gauge transformation and collecting the terms proportional to the equations of motion one finds 
 \begin{align}\label{delta_L1_nonloc}
   \delta_0\, \cL^1_{NL} \, &=  \, 
   \cE^{\a\b}\left\{\pr_{\mu}H_{\a\b}(\a^{\mu}+ \tal^{\mu}) + 
   H_{\a\nu}(\pr_{\b}\a^{\nu}-\pr^{\nu}\a_{\b})+
   H_{\mu\b}(\pr_{\a} \tal^{\mu}-\pr^{\mu} \tal_{\a})\right.\\
   &+ \, \left.\delta_0\big(H_{\a\b}\frac{\pr\cdot\cD}{\Box}\big)\right\}, 
 \end{align}
where
\be
\cE_{\m \n} \, = \, \Box \, H_{\m\n} \, - \, \partial_{\m} \, \partial^{\a} \, H_{\a}{}_{\n} \, -\, \partial_{\n} \, \partial^{\a} \, H_{\m}{}_{\a} \, 
+ \, \pr_{\m} \, \pr_{\n} \, \fr{\prd \prd H}{\Box} \, 
\ee
is the tensor defining the equations  stemming from \eqref{L=RR}. The variation \eqref{delta_L1_nonloc} will be relevant to compute the deformation of the gauge symmetry to first order in the nonlocal setup.

\subsection{Comparing  to the ${\cal N}=0$ supergravity Lagrangian} \label{sec:N=0}

Having led the cubic step of the Noether procedure to completion, we can now compare our result to the trilinear interactions of $\cN=0$ supergravity, whose action is
\begin{align}\label{N=0}
\mathcal{S}_{\mathcal{N}=0}=\int d^{D} x \sqrt{-g}\left\{\frac{1}{2 \kappa^{2}} R-\frac{1}{6} e^{-\frac{4 \kappa \varphi}{D-2}} \cH_{\mu \nu \lambda} \cH^{\mu \nu \lambda}-\frac{1}{2(D-2)} \partial_{\mu} \varphi \partial^{\mu} \varphi\right\}\, ,
\end{align} 
where $\kappa^2 = 8 \pi G^{(D)}$ is the $D-$dimensional gravitational coupling.

We already observed that our TT vertex \eqref{TTcubic} correctly reproduces the results of the DC for tree-level, three-point scattering amplitudes which, in their turn, can be derived from \eqref{N=0} too.  What we would like to stress here is that the matching extends off-shell. 

To this end, we focus on  \eqref{L_utc} where we employ the decomposition \eqref{hBphi}, writing $H_{\mu\nu}=h_{\mu\nu}+B_{\mu\nu}+\gamma\eta_{\mu\nu}\vf$.  Several vertices emerging from this substitution are proportional to the free equations of motion and as such can be field-redefined away to this 
order. For instance, purely scalar couplings in \eqref{L_loc} or in \eqref{L_utc} must be of the form 
\be
\sim \vf \, \pr \vf \, \pr \vf\, \sim \, \vf^2 \, \Box \vf \, .
\ee
In addition, performing the choice
\be \label{gamma}
 \g \, =\, \frac{1}{D-2}\, 
\ee  
the following two couplings get removed:
\be
\vf\,\partial^{\alpha} B_{\alpha \beta}\, \partial_{\mu} B^{\mu \beta}\, , \qquad
\vf\,\partial^{\alpha} h_{\alpha \beta}\, \partial_{\mu} h^{\mu \beta}\, .
\ee
Furthermore, the choice \eqref{gamma} of the value of $\g$, that was mentioned to play a special role already at the quadratic level, guarantees that the dilaton kinetic term as well as all the other cubic couplings involving the dilaton get normalised in agreement with \eqref{N=0}. Finally, let us recall that the action of $\mathcal{N}=0$ supergravity can be equivalently presented in the Einstein or in the string frame, the former corresponding to eq. \eqref{N=0}. We note that the definition \eqref{hBphi} of the graviton, with the value \eqref{gamma} of $\gamma$, corresponds to the linearisation of the Weyl rescaling connecting the two frames:
\begin{align}\label{Einstein_vs_String}
g^E_{\mu\nu}=e^{-\frac{2\,\kappa\,\varphi}{D-2}}\,g^S_{\mu\nu}\,.
\end{align}
Namely, if we let $g^E_{\mu\nu}=\h_{\mu\nu}+2\,\kappa\,h_{\mu\nu}$ and $g^{S}_{\mu\nu}=\h_{\mu\nu}+2\,\kappa\,H^S_{\mu\nu}$, then \eqref{Einstein_vs_String} is satisfied to first order in $\kappa$.

With these provisos, after some algebra, one finds indeed a correct match with the perturbative expansion of \eqref{N=0} to cubic order, including all longitudinal terms. In particular, as we already stressed, selecting only those couplings in $H_{\m\n}$ that respect the twofold Lorentz symmetry  one reproduces at the cubic level exactly the vertex coming from the expansion to first order in $\vf$ and zeroth order in $h_{\m\n}$ of the vertex involving the scalar field, the Kalb-Ramond field and the graviton:
\begin{gather} \label{eq:expphi}
-\frac{1}{6}\text{exp}\big[-\frac{4\kappa}{D-2}\vf\big] \cH_{\mu\nu\l}\cH_{\a\b\gamma}g^{\mu\a}g^{\nu\b}g^{\l\gamma}\, ,
\end{gather}
including the numerical coefficient that would be otherwise arbitrary under the conditions imposed by the Noether procedure alone.

 We can also compare our result to the cubic vertex of the Lagrangian obtained in \cite{Bern_Lagrangian}, where the DC field was defined in momentum space as the product of two YM fields deprived of their color indices. Consistently with the goal of reproducing amplitudes, in the gauge fixed construction of \cite{Bern_Lagrangian} all longitudinal terms  were discarded resulting in particular in a fully diagonal kinetic term of the form 
\be
\cL_{kin} \, = \, \12 \, H^{\m\n} \, \Box \, H_{\m\n} \, ,
\ee
where we are denoting still with $H_{\m\n}$ the resulting DC field. It is then simple to show that the DC cubic vertex resulting from the squaring procedure defined in \cite{Bern_Lagrangian} is equivalent to our  TT cubic vertex \eqref{TTcubic}. To reproduce higher order vertices one needs to also  integrate over the auxiliary fields introduced in \cite{Bern_Lagrangian}. The DC quartic vertex, in particular, was obtained in this fashion in \cite{pietro_tesi}. 

\section{Deformation of the gauge symmetry} \label{sec:deformation}

In this section we would like to compute the deformation of the gauge transformation of the DC field to first order in the coupling. The relevant equation 
is the second one in \eqref{Noether}, that we report here for convenience (in a schematic form with omitted indices):
\begin{align} \label{Noether1}
\frac{\d {\cal L}_0}{\d H} \,(\d_0^{\, (1)} H \, + \, \d_1^{\, (0)} H) \, + \, \frac{\d {\cal L}_1}{\d H} \,  \d_0^{\, (0)} H \, = \, 0 \, , 
\end{align}
where $\d_0^{\, (0)} H$ is the transformation \eqref{deltaH} with the parameter satisfying \eqref{tdiff}, $\d_0^{\, (1)} H$ is again given by  \eqref{deltaH}, but with parameter now subject to \eqref{corr1}, while $\d_1^{\, (0)} H$ is the first explicit nonlinear correction to the gauge transformation of the DC field.
Since our main interest is to explore the relation between geometry and DC, in this section we won't discuss explicitly the local theory with the auxiliary field $D$.

\subsection{First-order corrections to $\d_0 H_{\m\n}$} \label{sec:beyond_linear}

Let us consider first the local cubic theory defined in \eqref{L_loc}. By comparing equations \eqref{L_loc} and \eqref{deltaL_loc}, together with the gauge variation \eqref{varL0}, we can identify one solution to \eqref{Noether1} given by the following deformations of the Abelian symmetry \eqref{deltaH} and of the constraint \eqref{tdiff}:
\be \label{deform_local}
\begin{split}
\d_1^{(0)} H_{\,\m \n} \, &= \, (\a^{\g}+\tal^{\g})\,\pr_{\g}H_{\a\b}-H_{\a\g}\pr^{\g}\a_{\b}-H_{\g\b}\pr^{\g}\tal_{\a}+H_{\a\g}\pr_{\b}\a^{\g}+H_{\g\b}\pr_{\a}\tal^{\g}\, , \\
\pr^{\,\m } (\a_{\m} \, + \, \tilde{\a}_{\m}) &= \, - \frac{\lambda-2}{4}\, H^{\m\n}\, \left(\pr_{\mu}\, \a_{\nu} \, + \, \pr_{\nu} \, \tal_{\m}\right)\, .
\end{split}
\ee
Other solutions could be found upon employing equivalent forms for \eqref{deltaL_loc}. In particular, using the relation 
\be
\pr^{\m}\, M_{\m \n} \, = \, - \, \pr_{\n}\, \prd \prd H\, ,
\ee
expanding the second line in \eqref{deltaL_loc} and integrating by parts, we could generate new terms proportional  to the free equations depending on the parameter $\lambda$. Moreover, field redefinitions could modify the form of the cubic vertex in \eqref{L_loc} and consequently the deformations in \eqref{deform_local}.   One natural option would be to look for (a class of) deformations amenable to reproduce the Lie derivative on the fields of the DC multiplet to this order, so as establish a direct link with the underlying spacetime geometry. In the next section we shall provide an explicit analysis of this aspect. 

At any rate, it is not hard to appreciate that there is some tension between the DC origin of \eqref{L_loc} and the possibility to make the role of Riemannian geometry explicit. Indeed, one natural selection principle in the class defined by the possible field redefinitions of \eqref{deform_local} would be to ask for the second equation to reproduce the covariantised transversality condition, up to terms quadratic in the graviton field:
\be
\pr^{\m} \, (\a_{\m} + \tal_{\m}) + g\, \cO_1 (\a, \tal, H) \, = \nabla^{\m} \, (\a_{\m} + \tal_{\m}) + \cO(g^2) \, ,
\ee
where $\nabla^{\m}$ denotes the covariant derivative. Such a perturbative reconstruction would require to include the trace of the graviton, as can be seen expanding the Levi-Civita connection:
\be
\begin{split}
\nabla_{\m} \a^{\m} \, &= \, \left(\pr_{\m} \a_{\r} \, -  \, \G^{\s}_{\r \m} \a_{\s}\right) g^{\r \m}\, , \\
                                  &= \prd \a \, - \, \pr_{\m } \a_{\r} h^{\m\r} - \a^{\s} \left(\prd h_{\s} - \12 \pr_{\s} h^{\m}{}_{\m}\right) + \cO(h^2) \, ,
 \end{split}
\ee
but the presence of $h^{\m}{}_{\m}$, and then of $H^{\m}{}_{\m}$, would eventually spoil the left-right Lorentz symmetry implemented in the construction of the Lagrangian \eqref{L_loc}.

For the nonlocal case the relevant equations are \eqref{L_utc} and \eqref{delta_L1_nonloc}, while in \eqref{Noether1} there is no contribution to $\d_0^{(1)}$ to evaluate, since the gauge symmetry is now unconstrained.  The resulting deformation of the gauge transformation is 
\be \label{deform_nonloc}
\begin{split}
\d_1\,  H_{\,\m \n} \, = &\, \pr_{\r}H_{\m\n}(\a^{\r}+ \tal^{\r}) +  H_{\m\r}(\pr_{\n}\a^{\r}-\pr^{\r}\a_{\n})+
   H_{\r\n}(\pr_{\m} \tal^{\r}-\pr^{\r} \tal_{\m})\\
   &+ \delta_0\big(H_{\m\n}\frac{\pr\cdot\cD}{\Box}\big) \, ,
\end{split}
\ee
and displays a nonlocal correction to \eqref{deform_local} that involves longitudinal components of $H_{\m\n}$. 

\subsection{Covariance beyond the linear order} \label{sec:beyond_linear2}

In order to explore the relation between \eqref{deform_nonloc} and the Lie derivative of the fields entering the gravitational multiplet let us first notice that
 \be\label{observ}
   H_{\a\nu}(\pr_{\b}\a^{\nu}-\pr^{\nu}\a_{\b})+
   H_{\mu\b}(\pr_{\a} \tal^{\mu}-\pr^{\mu} \tal_{\a}) =
   H_{\a\nu}\pr_{\b}(\a^{\nu}+ \tal^{\nu})+
   H_{\nu\b}\pr_{\a}(\a^{\nu}+ \tal^{\nu})-
   \delta_0(H^{\a\nu}H_{\nu}{}^{\b})\, .
 \ee
Using \eqref{observ} and performing the field redefinition 
\begin{gather} \label{redef}
H_{\mu\nu}\rightarrow H_{\mu\nu}- H_{\mu\a}H^{\a}_{\;\;\nu}+\frac{1}{2} H_{\mu\nu}\frac{\pr\cdot\pr\cdot H}{\Box},
\end{gather}
allow one to recast the variation of the corresponding, redefined cubic vertex $\tilde{\cL}^1_{NL}$ as follows
 \begin{gather} \label{newvar}
   \delta\tilde{\cL}^1_{NL}= \cE^{\a\b}\big\{\pr_{\mu}H_{\a\b}(\a^{\mu}+ \tal^{\mu}) +
   H_{\a\nu}\pr_{\b}(\a^{\nu}+ \tal^{\nu})+
   H_{\nu\b}\pr_{\a}(\a^{\nu}+ \tal^{\nu})\big\},
 \end{gather}
from which one obtains the correction to the gauge transformation in the form
 \be \label{eq:delta1}
   \delta_1 
   H_{\a\b}=2\big\{\xi^{\nu}\pr_{\nu}H_{\a\b}+H_{\a\nu}\pr_{\b}\xi^{\nu}+
   H_{\nu\b}\pr_{\a}\xi^{\nu}\big\} \, ,
 \ee
where 
\be
\xi_{\m} \, = \, \12 \, (\a_{\m} \, + \, \tal_{\m}) \, ,
\ee
thus reproducing  the action of the Lie derivative on a symmetric tensor  and on a two-form $B_{\mu\nu}$ defined as in \eqref{hBphi}. 

Let us stress that, similarly to what observed at the end of the previous section,  the field redefinition \eqref{redef}, that we exploited in our attempt to recover the Lie derivative to cubic order, breaks the left-right Lorentz symmetry due to the off-diagonal contraction of indices in the second term.  In addition, \eqref{eq:delta1} can hardly suffice to establish full contact with the first nonlinear correction to diffeomorphisms, since it fails to reproduce the Lie derivative of the scalar field entering the gravitational multiplet defined in \eqref{hBphi}. Indeed, the correction to the (vanishing) free gauge transformation of $\vf$ stemming from \eqref{eq:delta1} is rather cumbersome and reads
\be 
\begin{split}
\delta_1\vf &=\xi\cdot\pr H + (\pr^{\mu}\xi^{\nu}+\pr^{\nu}\xi^{\mu})H^S_{\mu\nu}-\frac{1}{\Box}\big(\pr^{\mu}\pr^{\nu}\xi^{\a}\pr_{\a}H^S_{\mu\nu}+2(\pr^{\nu}\xi^{\a}+\pr^{\a}\xi^{\nu})\pr_{\a}\pr\cdot H^S_{\nu}\\ \label{eq:delta1f} &+\xi^{\a}\pr_{\a}\pr\cdot\pr\cdot H^S+2\pr^{\mu}\pr^{\nu}\xi^{\a}\pr_{\mu}H^S_{\a\nu}+2\Box\xi^{\a}\pr\cdot H^S_{\a}+\Box(\pr^{\mu}\xi^{\nu}+\pr^{\nu}\xi^{\mu})H^S_{\mu\nu}\big).
\end{split}
\ee
We interpret the mismatch as due to the attempt to derive nonlinear properties of the gauge transformation of the scalar field employing a definition of the latter that was tailored to the linear theory. In a complete theory, including all nonlinear contributions from $H_{\m\n}$, a  ``geometrical'' scalar,  should more likely be expressed as a combination of powers of $H_{\m\n}$ of increasing order:
\begin{gather} 
\psi = \psi^{(1)}+\psi^{(2)}+\psi^{(3)}+...,
\end{gather}
with $\psi^{(n)} \sim O (H^n)$ and $\psi^{(1)}=\vf$ as defined in the third of \eqref{hBphi}. In order to get a clue on the systematics of these prospective nonlinear corrections, let us observe that  our definition \eqref{hBphi} of the scalar field can be 
formally read as:
\begin{gather}  \label{phi_geom}
\vf = -\frac{R^{(1)}}{\Box}\, ,
\end{gather}
where $R^{(1)}$ is the linearised Ricci tensor for $H^S_{\m\n}$. Given that we are after a scalar under diffeomorphisms, we are naturally led to a geometrical guess for the general solution, in the form
\begin{gather} \label{eq:fullpsi}
\psi := -\frac{R}{\hat{\Box}},
\end{gather}
where $R = R^{(1)} + R^{(2)} + \ldots  $ is the Ricci scalar of the metric $g_{\mu\nu}=\eta_{\mu\nu}+H^S_{\mu\nu}$. One might expect that the metric of the actual spacetime involve $h_{\m\n}$
rather than $H^S_{\mu\nu}$. However, as noticed in \eqref{Einstein_vs_String}, at least at the linearised level the two are connected by the Weyl rescaling connecting Einstein and string frame in $\mathcal{N}=0$ supergravity. Here we choose to carry on this exercise in terms of $H^S_{\mu\nu}$, as it conveys anyway an idea of the involved relation between DC and geometry. Expanding the Ricci scalar perturbatively, we have
\begin{align} 
R^{(1)}&=\pr\cdot\pr\cdot H^S - \Box H^S,\\ \nn
R^{(2)}&=H_S^{\mu\nu}\Box H^S_{\mu\nu}-
\12\pr^{\nu}H_S^{\mu\a}\pr_{\a}H^S_{\mu\nu}+
\frac{3}{4}\pr^{\a}H_S^{\mu\nu}\pr_{\a}H^S_{\mu\nu}-2H^{\mu\nu}_S\pr_{\mu}\pr\cdot H^S_{\nu}\\  \label{eq:ricci2} &+
H_S^{\mu\nu}\pr_{\mu}\pr_{\nu}H^S-
\pr\cdot H_S^{\a}\pr\cdot H^S_{\a}+
\pr^{\a}H^S \pr\cdot H^S_{\a}-
\frac{1}{4}\pr^{\a}H^S\pr_{\a}H^S\, ,
\end{align}
while $\hat{\Box}$ is the Laplace-Beltrami operator for a scalar field, \textit{i.e.}:
\begin{gather} 
\hat{\Box}:=g^{\mu\nu}(\pr_{\mu}\pr_{\nu}-\Gamma^{\a}_{\mu\nu}\pr_{\a}),
\end{gather}
that can be also expanded in powers of $H^S_{\mu\nu}$:
\begin{gather} 
\hat{\Box}^{(0)}=\eta^{\mu\nu}\pr_{\mu}\pr_{\nu}=\Box,\\
\label{eq:hatbox1}\hat{\Box}^{(1)}=-H_S^{\mu\nu}\pr_{\mu}\pr_{\nu}-\pr\cdot H_S^{\a}\pr_{\a}+\12 \pr^{\a}H^S\pr_{\a}.
\end{gather}
In particular, using the second of \eqref{eq:ricci2} and  \eqref{eq:hatbox1} we can compute the conjectured form of the first correction to $\vf$ implied in \eqref{eq:fullpsi}:
\begin{align} \label{psi2}
\begin{split}
\psi^{(2)}=&-\frac{1}{\Box}\Big(H_S^{\mu\nu}\Box H^S_{\mu\nu}-
\12\pr^{\nu}H_S^{\mu\a}\pr_{\a}H^S_{\mu\nu}+
\frac{3}{4}\pr^{\a}H_S^{\mu\nu}\pr_{\a}H^S_{\mu\nu}-2H^{\mu\nu}_S\pr_{\mu}\pr\cdot H^S_{\nu}\\  &+
H_S^{\mu\nu}\pr_{\mu}\pr_{\nu}H^S-
\pr\cdot H_S^{\a}\pr\cdot H^S_{\a}+
\pr^{\a}H^S \pr\cdot H^S_{\a}-
\frac{1}{4}\pr^{\a}H^S\pr_{\a}H^S \\
&-H_S^{\mu\nu}\pr_{\mu}\pr_{\nu}H^S+
H_S^{\mu\nu}\pr_{\mu}\pr_{\nu}\frac{\pr\cdot\pr\cdot H^S}{\Box}
-\pr\cdot H_S^{\a}\pr_{\a}H^S\\ &+
\pr\cdot H_S^{\a}\pr_{\a}\frac{\pr\cdot\pr\cdot H^S}{\Box} +
\12 \pr^{\a}H^S\pr_{\a}H^S-
\12 \pr^{\a}H^S\pr_{\a}\frac{\pr\cdot\pr\cdot H^S}{\Box}\big).
\end{split}
\end{align}
As a test of our guess we consider the transformation properties of $\psi$, order by order,
\begin{gather} \label{eq:deltapsifull}
\delta\psi = \xi\cdot\pr\psi \Rightarrow
\begin{cases}
\delta_0\vf=0,\\
\delta_1\vf+\delta_0\psi^{(2)}=\xi\cdot\pr\vf,\\
\delta_2\vf + \delta_1\psi^{(2)}+\delta_0\psi^{(3)}=\xi\cdot\pr\psi^{(2)},\\
...
\end{cases}
\end{gather}
and use the form of $\delta_1\vf$ computed in \eqref{eq:delta1f} in order to determine $\d_0 \psi^{(2)}$. After some algebra we find
\begin{align} \nn
\delta_0\psi^{(2)}&=\frac{1}{\Box}\delta_0\Big(
-H_S^{\mu\nu}\Box H^S_{\mu\nu}+
\12\pr^{\nu}H_S^{\mu\a}\pr_{\a}H^S_{\mu\nu}-
\frac{3}{4}\pr^{\a}H_S^{\mu\nu}\pr_{\a}H^S_{\mu\nu}+2H^{\mu\nu}_S\pr_{\mu}\pr\cdot H^S_{\nu}\\ \nn &-
H_S^{\mu\nu}\pr_{\mu}\pr_{\nu}\frac{\pr\cdot\pr\cdot H^S}{\Box}+
\pr\cdot H_S^{\a}\pr\cdot H^S_{\a}-
\pr^{\a}\frac{\pr\cdot\pr\cdot H^S}{\Box}\pr\cdot H^S_{\a}\\ \label{eq:delta0psi2,2} &+
\frac{1}{4}\pr^{\a}\frac{\pr\cdot\pr\cdot H^S}{\Box}\pr_{\a}\frac{\pr\cdot\pr\cdot H^S}{\Box} \Big)\, ,
\end{align}
that can be easily checked to match the corresponding variation of \eqref{psi2}, to this order.

\section{Outlook} \label{sec: outlook}

Our exploration of the DC shows that the correspondence between spin-one gauge theories and gravity extends quite neatly off shell at the quadratic level, including
the longitudinal sectors of the corresponding Lagrangians. Both at the quadratic and at the cubic level it is also clear that there is more to the DC than just a perturbative implementation of gauge invariance, that by itself would not suffice neither to fix the free Lagrangian in the form \eqref{L=RR} nor to recover uniquely the
cubic couplings of the $\cN=0$ supergravity action \eqref{N=0}. There emerge also concrete indications that having Riemannian geometry explicitly implemented seems to hide the underlying DC structure, in particular in view of the non-polynomial field redefinitions likely to be needed in order to recover the Lie derivative of the various physical fields. It might be that a more appropriate context to assess the geometrical meaning of the DC has to be found in  double field theory \cite{Hull:2009mi}, as  explored in \cite{Lee:2018gxc,Cho:2019cl,Kim:2019jwm,Lescano:2020nve}.

One of the aspects that would be interesting to investigate further is the option of resumming all perturbative corrections so as to make sense at the nonlinear level of an action of the form \eqref{L=RR}, written as a square of curvatures. It is legitimate to speculate that the double-copy meaning of such an action may be related to the effective Yang-Mills Lagrangian of \cite{Bern_Lagrangian, Tolotti:2013caa}. It would be also relevant to understand how much of the simplicity of the massive deformation of the free theory  \eqref{L=RR+m} may be kept at the interacting level, so as to make more concrete contact with the various incarnations of the DC for massive gravity, in the spirit of  \cite{Johnson:2020pny,Momeni:2020vvr}. See also  \cite{Chiodaroli:2015rdg,Chiodaroli:2017ehv,Chiodaroli:2018dbu}.

A better understanding of the DC at the Lagrangian level should contribute to clarify its status at the level of classical solutions \cite{BH_and_DC,Luna:2015paa,Luna:2016hge,Ridgway:2015fdl,Bahjat-Abbas:2017htu,Carrillo-Gonzalez:2017iyj,Plefka:2018dpa,Luna:2018dpt,Gurses:2018ckx,Andrzejewski:2019hub,Sabharwal:2019ngs,PV:2019uuv}. In particular, it might  help to improve the calculational schemes for gravitational radiative solutions viewed as squares of solutions describing gluonic radiation \cite{Luna:2016due,Goldberger:2016iau,Goldberger:2017vcg,Chester:2017vcz,Li:2018qap,Shen:2018ebu,Bern:2019nnu,Bern:2019crd}, for instance by allowing one to compare different options for gauge fixings or frame choices. Furthermore, upon properly identifying the boundary conditions on the two sides of the DC it should be possible to explore the relations between the corresponding asymptotic symmetries \cite{strominger_review}, thus providing an alternative perspective on soft theorems and memory effects \cite{Oxburgh:2012zr,Vera:2014tda,DiVecchia:2017gfi,Bautista:2019tdr,A:2020lub}. 

The existence of DC relations points to a deeper meaning of the notion of local symmetry. It would be interesting to explore its possible implementation for higher-spin gauge theories. On the one hand, the three quadratic DC  Lagrangians that we presented in this work, \eqref{DClagrML}, \eqref{tripletnosymm} and \eqref{L=RR}, actually are low-spin representatives  of an infinite class of theories involving gauge fields of arbitrary spins stemming from tensionless strings, and it is at least conceivable that all the members of those infinite classes might be interpreted along similar lines. Furthermore, concrete indications of double-copy (and multiple-copy) type structures at the interacting level have been noticed for higher-spin vertices \cite{Manvelyan:2010je, Taronna:2011kt}, thus suggesting the existence of general structures that would deserve further investigation.

\acknowledgments

P.F. would like to thank I. Basile for discussions. D.F. gratefully acknowledges exchanges with K. Mkrtchyan and S. Nagy. D.F. is grateful to the Asian-Pacific Center for Theoretical Physics (APCTP) in Pohang for the kind hospitality extended to him during the preparation of this work.

\begin{appendix}

\section{Notation and conventions}\label{sec:notation}

We adopt the ``mostly plus'' metric convention for the Minkowski metric,
$
\eta_{\mu\nu}=\text{diag}(-1,+1,...,+1),
$
and the following sign convention for the Riemann tensor (without assuming a torsion-free connection): 
\begin{align}
 R^{\rho}_{\hspace{1.5mm}\sigma\mu\nu}=\pr_{\mu}\Gamma^{\rho}_{\nu\sigma}-
\pr_{\nu}\Gamma^{\rho}_{\mu\sigma}+
\Gamma^{\rho}_{\mu\l}\Gamma^{\l}_{\nu\sigma}-
\Gamma^{\rho}_{\nu\l}\Gamma^{\l}_{\mu\sigma}.
\end{align}
Whenever there are no ambiguities due to the position of indices, we may employ the shorthands $\pr\cdot$ for a divergence, as in $\prd \prd H :=\pr^{\a} \pr^{\b} H_{\a \b}$, and  $H := H^{\a}_{\;\a}$ for the trace.

We denote the Fourier transform with the symbol $\cF$, with normalisation and signs defined as follows:
\begin{align}
f(x) =\int \frac{d^Dp}{(2\pi)^D}\, e^{-ip\cdot x} \cF\{f\}(p)\, , \qquad 
 \cF\{f\}(p)&=\int d^Dx\, e^{ip\cdot x} f(x)\, ,
\end{align}
and employ  the symbol $\circ$ for the convolution product of two functions:
\be  \label{convolution}
[f\circ g](x)= \int d^Dy f(y)g(x-y).
\ee
This product is commutative and associative, and {\it does not} satisfy the Leibniz rule. Rather,
\be \label{nonLei}
\pr_{\m} \, (f\circ g) \, = \, (\pr_{\m} \, f) \, \circ g \, = \, f \circ (\pr_{\m} \, g) \, .
\ee
%

\section{Twofold Lorentz symmetry}\label{appendixB}

Given that, in the DC scenario, gravity amplitudes are to be obtained from the product of two factors  which are separately Lorentz invariant, it is natural to expect that such a twofold Lorentz symmetry be mirrored, somehow, at the Lagrangian level. A field-theoretical linear realisation of invariance under $O(D-1,1)_L\otimes O(D-1,1)_R$ was first obtained in the low-energy effective action of the closed string  in \cite{Siegel:1993th,Siegel:1993bj,Siegel:1993xq}, and later identified in the context of Double Field Theory  as well \cite{Hull:2009mi}. In \cite{Bern:1999ji}, a gravity action whose vertices respect this symmetry was built up to the quintic order and in \cite{Cheung:2016say} it was shown the existence of local field redefinitions that make this symmetry manifest to all orders in a gauge-fixed version of the Einstein-Hilbert action. 

In our context, it is precisely such twofold Lorentz symmetry that allows us to identify uniquely the quadratic Lagrangian \eqref{Lnonloc}, as we shall now show. A similar argument also applies to the TT cubic vertex \eqref{TTcubic}. Let us begin by considering the most general quadratic form in $H_{\mu\nu}$ involving two derivatives:
\be
 \begin{split} \label{free}
  \cL^{(0)}_{NL} \, = \, & a_1 \, \pr_{\m}\, H_{\a\b} \, \pr^{\m} \, H^{\a\b}\, +\, a_2\, \pr_{\m}H_{\a\b} \, \pr^{\m} \, H^{\b\a} \, + \, 
   a_3\, \pr^{\a} \, H_{\a\b} \, \pr_{\g} \, H^{\g \b} \\
   & + \, a_4 \, \pr^{\a}\, H_{\a\b}\, \pr_{\g} \, H^{\b\g} \, + \, a_5\, \pr^{\a}\, H_{\b\a}\, \pr_{\g}\, H^{\b\g} \, + \, a_6\, \pr^{\a}\, H_{\a\b}\, \pr^{\b}\, H \\
   & + \, a_7\, \pr_{\a}\, H\, \pr^{\a}\, H \, + \, a_8 \, \prd \prd H\, \fr{1}{\Box}\, \prd \prd H\, ,
 \end{split} 
\ee
where the nonlocal term accounts for the projector appearing in \eqref{hBphi}.  Upon imposing invariance of \eqref{free} under \eqref{deltaH} 
one finds the general solution
\begin{align}
\begin{split}
  \cL^{(0)}_{NL}\, = &\,  
   a \, \pr_{\mu}H_{\a\b}\pr^{\mu}H^{\a\b}  +
   b \, \pr_{\mu}H_{\a\b}\pr^{\mu}H^{\b\a}  - a \, \pr^{\a}H_{\a\b}\pr_{\gamma}H^{\gamma\b}  - 
   2b \, \pr^{\a}H_{\a\b}\pr_{\gamma} \, H^{\b\gamma} \\
  -&   a\pr^{\a}H_{\b\a}\pr_{\gamma}H^{\b\gamma} 
 + \,  2(a+b+c) \, \pr^{\a}H_{\a\b}\pr_{\b}H \, -
 \, (a+b+c) \, \pr_{\a}H\pr^{\a}H\, \\
 +& \,c\, \prd \prd H\, \fr{1}{\Box}\, \prd \prd H\,,
\end{split}
\end{align}
depending in principle on two arbitrary parameters, up to the overall normalisation. Imposing an additional symmetry where each of the two indices in $H_{\m\n}$ undergoes a separate Lorentz transformation (while derivatives transform according to the index they are contracted with), leads to select the values $b = 0$ and  $c = + \tfrac{1}{2}$, while  $a = - \tfrac{1}{2}$ results from the request of canonical normalisation of the kinetic term. These choices select the Lagrangian \eqref{L=RR},
\begin{align} \label{RRbis}
 \cL_{NL} \, = \, \12 H^{\a\b}\big\{ \Box \eta_{\a\mu}\eta_{\b\nu}-\pr_{\a}\pr_{\mu}\eta_{\b\nu}
  -\pr_{\b}\pr_{\nu}\eta_{\a\mu}+\pr_{\a}\pr_{\b}\frac{1}{\Box} \pr_{\mu}\pr_{\nu}
  \big\}H^{\mu\nu} \, ,
\end{align}  
whose  kinetic operator  can be seen as the product of two Maxwell operators with the insertion of an inverse d'Alembertian:
 \be 
   \big\{ \Box \eta_{\a\mu}\eta_{\b\nu}-\pr_{\a}\pr_{\mu}\eta_{\b\nu}
  -\pr_{\b}\pr_{\nu}\eta_{\a\mu}+\frac{\pr_{\a}\pr_{\b}\pr_{\mu}\pr_{\nu}}{\Box} 
  \big\} = \big\{\Box \eta_{\a\mu}-\pr_{\a}\pr_{\mu}  
   \big\}\frac{1}{\Box}\big\{\Box \eta_{\b\nu}-\pr_{\b}\pr_{\nu}
   \big\}\, .
 \ee
The local counterpart of the above Lagrangian, given by \eqref{DClagrML}, can be viewed either as a consistent truncation of \eqref{RRbis} to its sector with only transverse local symmetry \eqref{tdiff}, or, again, as the unique {\it local} solution that respects the twofold Lorentz symmetry, and thus in particular does not include  terms proportional to traces.  If the twofold Lorentz symmetry were discarded and traces were included in a theory without the transversality condition \eqref{tdiff}, the solution would be a mixed-symmetry Lagrangian of the type described in \cite{labastida}, whose equations of motion would still propagate a graviton and a two-form, but not a scalar. 

\section{The Noether procedure}\label{noetherAlg}

The following illustration of the Noether method is based on \cite{ML_cubic}. Let us suppose that both the action functional $S [\vf]$ and its gauge invariances $\d  \vf$  admit a perturbative expansion in powers of the fields:
\be 
\begin{split}
& S \, [\vf] \, = \, S_0 \, [\vf] \, + \, g\, S_1 \, [\vf] \, + \, g^{\,2}\, S_2 \, [\vf] \, \ldots \, ,\\ 
& \d \, \vf \, = \, \d_0 \, \vf \, + \, g \, \d_1 \, \vf  \, + \, g^{\, 2} \, \d_2 \, \vf  \, \ldots \, ,
\end{split}
\ee
where  $\vf$ collectively denotes all types of fields entering the theory, while the coupling $g$ plays the role of a counting parameter. $S_0 \, [\vf]$ is the free, quadratic action  while  $S_k \, [\vf] \sim \vf^{\, k + 2}$. Similarly, $\d_0 \, \vf$ is the Abelian gauge symmetry of the free action, $\d_0 \, \vf = \pr \, \e$,  while the higher order contributions  are linear in the infinitesimal gauge parameter and depend in principle on an increasing number of fields: $\d_k \vf  \sim \, \e \, \vf^{\, k}$. Perturbative gauge invariance of the action holds if all the equations of the Noether system are satisfied:
\be \label{noether}
\begin{split}
& \d_0 \, S_0 \, [\vf] \, = \, 0 \, ,\\ 
& \d_1 \, S_0 \, [\vf] \, + \, \d_0 \, S_1 \, [\vf] \, = \, 0\, ,\\
& \d_2 \, S_0 \, [\vf] \, + \, \d_1 \, S_1 \, [\vf] \, + \, \d_0 \, S_2 \, [\vf] \,= \, 0\, ,\\
& \dots \, .
\end{split}
\ee
Their solutions provide the possible interactions compatible with gauge invariance, while at the same time determining the corresponding deformations, Abelian or non-Abelian, of the free gauge symmetry.  

As it happens for \eqref{DClagrML}, the free gauge parameters may be subject to given off-shell conditions implemented through the action of some (linear) operator ${\cal O}$ in the schematic form
\be \label{constraints}
{\cal O} \e \, = \, 0,
\ee
a concrete example being provided by the transversality condition \eqref{tdiff}.  In this case, the constraints \eqref{constraints}  may  themselves receive perturbative corrections in increasing powers of the fields:
\be \label{corr}
{\cal O} \e \, + \, g \, {\cal O}_1 \, (\e\, ,\vf) \, + \, g^2 \, {\cal O}_2 (\e\, , \vf^{\, 2}) \, + \, \ldots\, = \, 0 \, .
\ee
In order to properly take such corrections  into account one has to better specify the Noether equations \eqref{noether}. In particular, each of the terms $\d_k \, \vf$,  would admit in its turn a perturbative expansion in powers of $\vf$, due to their {\it implicit} dependence on $\vf$ hidden in $\e$ because of \eqref{corr}:
\be
\begin{split}
&\d_k \, \vf \, = \, \d_k^{\, (0)} \vf \, + \,  \d_k^{\, (1)} \vf \, + \, \d_k^{\, (2)} \, \vf \, + \, \ldots \, , \\
& \d_k^{\, (l)} \vf \, := O\, (\vf^{\, k+l}) \,
\begin{cases}
&\mbox{{\it explicitly} on} \sim \vf^{\, k},    \\
&\mbox{{\it implicitly} on} \sim \vf^{\, l} ,\, \mbox{via its $\e-$dependence.}
\end{cases}
\end{split}
\ee
Correspondingly,  the system \eqref{noether} gets modified as follows:
\begin{align} \label{Noether}
 o\, (\e, \vf)&: \, \frac{\d {\cal L}_0}{\d \vf} \,  \d_0^{\, (0)} \vf \, = \, 0 \, ,\nonumber \\ 
o\, (\e, \vf^{\, 2})&:   \, \frac{\d {\cal L}_0}{\d \vf} \,(\d_0^{\, (1)} \vf \, + \, \d_1^{\, (0)} \vf) \, + \,
\frac{\d {\cal L}_1}{\d \vf} \,  \d_0^{\, (0)} \vf \, = \, 0 \, , \\
o\, (\e, \vf^{\, 3})&:   \d \, \frac{\d {\cal L}_0}{\d \vf} \,(\d_0^{\, (2)} \vf  +  \d_1^{\, (1)} \vf +  \d_2^{\, (0)} \vf ) + 
\frac{\d {\cal L}_1}{\d \vf}  (\d_0^{\, (1)} \vf  +  \d_1^{\, (0)} \vf ) +  \frac{\d {\cal L}_2}{\d \vf}  \d_0^{\, (0)} \vf \, = \, 0 \, ,  \nonumber \\
\dots \, . \nonumber
\end{align}
For instance, in the case under scrutiny in this paper, both $\d_0^{\, (0)} H_{\m\n}$ and  $\d_0^{\, (1)} H_{\m\n}$ are given by the Abelian transformation  
$\pr_{\mu}\, \a_{\nu} \, + \, \pr_{\nu} \, \tal_{\m}$. However, the parameters in $\d_0^{\, (0)} H_{\m\n}$ solve $\pr^{\, \m} (\a_{\m} + \tal_{\, \m}) = 0$, while in  $\d_0^{\, (1)} H_{\m\n}$, they are the solution to
\be \label{corr1}
\pr^{\m} \, (\a_{\m} + \tal_{\m}) + g\, \cO_1 (\a, \tal, H) \, = \, 0 \, ,
\ee
thus carrying a possible correction to first order in $H_{\m\n}$. Similarly for the higher-order terms\ft{An alternative option in our case  would be to solve for the transversality condition \eqref{tdiff} in terms of unconstrained gauge parameters \cite{PopeStelle, AlvarezValea,FLS}, to then proceed with the implementation of the Noether method in the standard fashion. However, the corresponding high-derivative and reducible gauge symmetry would introduce complications of its own in the solution to \eqref{noether}. See \cite{unfree1, unfree2} for the Hamiltonian treatment of gauge systems with differentially constrained gauge symmetry.}. 

In this paper we employ the Noether procedure in order to find the cubic vertices deforming the three Lagrangians \eqref{DClagrML}, \eqref{tripletnosymm} and \eqref{Lnonloc}. While for the second and the third one the Abelian symmetry is unconstrained and we implement the procedure in its customary form \eqref{noether}, for \eqref{DClagrML} we need to consider the possibility of non-trivial deformations of \eqref{tdiff} of the form \eqref{corr}. For all cases, from the second equation of \eqref{Noether} one first determines the cubic vertices $\cL_1$ on the free mass shell, {\it i.e.} for those configurations that vanish when $ \tfrac{\d {\cal L}_0}{\d \vf}  =  0$ holds. At this stage the possible corrections encoded in \eqref{corr} do not play any role. Once $\cL_1$ is found one should collect all terms in $\d \cS$ that are quadratic in $\vf$  and that vanish on the free mass shell, and arrange them as in the second equation of \eqref{noether} or \eqref{Noether} so as to compute $\d_1 \vf$. 

In some cases with constrained gauge invariance, taking the corrections \eqref{corr} into account may be unavoidable in order to grant for the existence of a {\it local } solution to the deformation procedure \cite{ML_cubic}. In our very case, as we saw, while not strictly needed to the purpose of solving for the cubic vertices, taking the option \eqref{corr} into account can still be relevant: on the one hand, it may entail some algebraic simplifications, on the other hand, and more importantly, it may be connected with the underlying geometry of the theory.  For instance, the cubic vertices of unimodular gravity can be consistently computed without any need to take \eqref{corr} into account. Still, corrections as those encoded in \eqref{corr} are instrumental in that context to promote the transversality condition $\prd \e = 0$ to its covariant version $\nabla \cdot \e = 0$.  The same observation applies indeed to \eqref{tdiff}, but in the latter case recovering Riemannian geometry leads to obscure
the underlying DC structure.

\end{appendix}

\end{fmffile}
\end{document}